%% file: counterrotation_arxiv.tex
\documentclass{aa}  

\usepackage[colorlinks,citecolor=blue]{hyperref}
\usepackage{graphicx}
\usepackage{booktabs}
\usepackage{multirow}
\usepackage{enumitem}
\usepackage{color}

\input{notes.tex}

\AtBeginDocument{}%
\begin{document}

\title{Disk heating and bending instability in galaxies with counterrotation}
\titlerunning{Disk heating in galaxies with counterrotation}

\author{Sergey Khoperskov\inst{1,2} \and
Giuseppe Bertin\inst{3}}
\authorrunning{S. Khoperskov and
G.Bertin}

\institute{GEPI, Observatoire de Paris, CNRS, Universit\'e Paris Diderot, 5 place Jules Janssen, 92190 Meudon, France \\
\email{sergey.khoperskov@obspm.fr}
\and
Institute of Astronomy, Russian Academy of Sciences, 48 Pyatnitskaya st., 119017 Moscow, Russia
\and
Dipartimento di Fisica, Universit\`{a} degli Studi di 	Milano, via Celoria 16, I-20133 Milano, Italy\\
\email{giuseppe.bertin@unimi.it}
}

 
\abstract{With the help of high-resolution long-slit and integral-field spectroscopy observations, the number of confirmed cases of galaxies with counterrotation is increasing rapidly. The evolution of such counterrotating galaxies remains far from being well understood. In this paper we study the dynamics of counterrotating collisionless stellar disks by means of $N$-body simulations. We show that, in the presence of counterrotation, an otherwise gravitationally stable disk can naturally generate bending waves accompanied by strong disk heating across the disk plane, that is in the vertical direction. Such conclusion is found to hold even for dynamically warm systems with typical values of the initial vertical-to-radial velocity dispersion ratio $\czcr \approx 0.5$, for which the role of pressure anisotropy should be unimportant. We note that, during evolution, the $\czcr$ ratio tends to rise up to values close to unity in the case of locally Jeans-stable disks, whereas in disks that are initially Jeans-unstable it may reach even higher values, especially in the innermost regions. This unusual behavior of the $\czcr$ ratio in galaxies with counterrotation appears not to have been noticed earlier. Our investigations of systems made of two counterrotating components with different mass-ratios suggest that even apparently normal disk galaxies (i.e., with a minor counterrotating component so as to escape detection in current observations) might be subject to significant disk heating especially in the vertical direction.}


   \maketitle
%

\section{Introduction}\label{sec::intro}

In the early stages of galactic evolution, dwarf galaxy mergers, tidal interactions with satellites, and gas accretion from
cosmological filaments lead to the formation of giant
disk galaxies made of stellar and/or gaseous subsystems with different angular momentum axes~\citep{2008MNRAS.386..909C,2011MNRAS.417..882D,2011MNRAS.417.2982F,2012ApJ...747...34B,2014MNRAS.437.3596A}. In such picture, disk galaxies can be thought of as multi-spin systems~\citep{1994AJ....108..456R,2014ASPC..486...85C}. In some cases, rotation occurs in distinct planes or even in perpendicular planes, in configurations that sometimes contain structures that are usually called inner or outer polar rings~\citep{1987ApJ...314..439W,2011MNRAS.418..244M,2012AstBu..67..147M}. These systems are interesting laboratories for the investigation of the dynamics of truly 3D matter distributions around galaxies~\citep{1994ApJ...436..629S,1996A&A...305..763C,2014MNRAS.441.2650K}. Galaxies with external elongated, but inclined subsystems are rare because their dynamics would make them evolve away from such conditions~\citep{1996A&A...305..763C}. The processes that lead to the formation of galaxies with subsystems rotating in the same plane are still under debate~\citep[][ and references therein]{1999IAUS..186..149B,2014ASPC..486...51C}. It is generally believed that inclined gaseous components tend to settle into the main galactic plane, thus forming corotating or counterrotating systems~\citep{1996ApJ...461...55T}. The possibility of accretion of gas and/or stars right onto the plane of the host galaxy is not to be excluded.

On the observational side, several galaxies with strong counterrotation have been studied in detail~\citep[e.g., see][]{1992ApJ...394L...9R,2009ApJ...694.1550S,2011BaltA..20..453K,2013ApJ...769..105K,2015A&A...581A..65C}. In turn, from the theoretical point of view, the evolutionary scenarios mentioned above suggest that such objects should not be so rare~\citep{2014MNRAS.437.3596A}. It is expected that disk galaxies of various morphological types can host a counterrotating component. The known list of confirmed cases of systems with counterrotation is often stated to contain about ten galaxies. This number is bound to increase rapidly, because of recent progress in integral-field and deep long-slit  spectroscopic observations~\citep{2011MNRAS.412L.113C,2014A&A...570A..79P,2015A&A...581A..65C,2016arXiv160504795B}. 

The main issues that justify the great interest in galaxies characterized by counterrotation are the following:
\begin{itemize}
\item the origin of galaxies with counterrotation: processes of accretion of gas and/or stars;
\item the observed physical properties of the counterrotating components (spatial distribution, kinematics, relative mass, stellar ages, etc.);
\item dynamical and secular evolution of galaxies with counterrotation.
\end{itemize}  
In this paper we focus primarily on the dynamical evolution of collisionless stellar systems with counterrotation under different initial conditions.

By means of a linear WKB analysis of a fluid model~\citep{2006A&A...459..333B}, it has been shown that non-axisymmetric perturbations arise in a galactic disk when two components are rotating with different velocities, and, in particular, when the two components are in counterrotation. For a disk made of two counterrotating components, another WKB analysis showed \citep{1997ApJ...475...83L} that the most unstable perturbations are one-armed leading waves (with respect to the rotation direction of the heavier component, assumed to be stellar) and that, in general, one-armed instabilities are more important than two-armed instabilities~\citep[see also][for nearly Keplerian discs]{2016arXiv160207169G}). Then, by means of 2-D simulations~\citep{1997ApJ...484L..33C} it was shown that collisionless counterrotating disks, initially characterized by an axisymmetric Jeans stability parameter above unity, are subject to the formation of one-armed patterns.

Obviously, the evolution of galaxies with counterrotation is expected to depend on the parameter regime associated with the structure of the disk. Specific ways to interpret the formation of counterrotating bars~\citep{1996A&A...312..761F,2006MNRAS.371..451M}, leading spiral structures~\citep{1993PASJ...45L..47V}, and other phenomena have been proposed and related to the parameter regimes that are involved. In general, it is recognized that the free energy offered by the presence of counterrotation is substantially the same as that at the basis of the so-called ``two-stream instability," known and studied in a variety of contexts~\citep{1971Ap&SS..14...52K,1971ApL.....9...37M,1976ApJ...206..418M,1980A&A....88..289B,1982A&A...106..274B,1987AJ.....94...99A}. By means of N-body simulations ~\citet{1994ApJ...425..530S} showed the existence of  bending modes directly related to the presence of counterrotation; in their models, various structures arise dependent on the initial pressure anisotropy conditions. Counterrotation appears to drive the bending instability across the disk plane in a way that is similar to the driving of the Kelvin-Helmholtz instability in collisional systems in the presence of shear flows. Recently~\citet{2015MNRAS.446..622Q} found that a supersonic Kelvin-Helmholtz instability can originate in the case of spatially separate counterrotating components in contact along an annular region, for which the instability is absent if the perturbation is required not to develop in the vertical direction; this suggests that one very important characteristic of the instability of counterrotating disks is its three-dimensional character. Therefore, in this paper we decided to focus, as our primary objective, on the dynamical evolution of counterrotating galaxies across the galactic plane.
 
Many important questions related to the evolution of counterrotating stellar disks have been addressed by~\citet{1994ApJ...425..530S}. In particular, thicker disks are expected to be more stable; because evolution usually proceeds in the direction of more stable configurations, it is natural to expect an increase of the vertical velocity dispersion during the evolution of unstable counterrotating disk. However, it appears that a direct quantitative study of the evolution of the vertical-to-radial velocity dispersion ratio $\czcr$ for systems with counterrotation has not been carried out so far. 

The $\czcr$ ratio plays an important role in the disk-halo decomposition of the kinematics of external galaxies \citep[e.g., see {\it The Disk Mass Project}; ][]{2004AN....325..151V, 2011ApJ...739L..47B}, because it affects the line-of-sight velocity dispersion, whereas on the plane the $\sigma_{\varphi}/\sigma_{\rm R}$ ratio is expected to be well constrained by the epicyclic theory.
In other words, knowledge or clues about the $\czcr$ ratio are important for the general problem of measuring the amount and distribution of dark matter in spiral galaxies. 

Commonly used values for the $\czcr$ ratio vary in the range $0.4-0.7$~\citep{1999A&A...352..129V}. There is some growing empirical evidence that the $\czcr$ ratio varies monotonically with Hubble type. Namely, $\czcr \approx 0.8$ for Sa-type galaxies and decreases down to $\approx 0.2$ for Scd galaxies~\citep{2003AJ....126.2707S,2012MNRAS.423.2726G}. This relation has been argued to result from the impact of various mechanisms of stellar disk heating, related to: stochastic spiral patterns~\citep{1990MNRAS.245..305J,2006MNRAS.368..623M},  bar structures~\citep{2010ApJ...721.1878S}, molecular cloud relaxation~\citep{1951ApJ...114..385S,1984MNRAS.208..687L,2016MNRAS.462.1697A}, disk-halo interaction~\citep{2001ApJ...563L...1F},  interactions with dark halo objects~\citep{2004A&A...421.1001H} or with black holes~\citep{1985ApJ...299..633L}, heating by infalling satellites~\citep{2004MNRAS.351.1215B}, and other processes~\citep{2011ARA&A..49..301V}.

Several studies of the dynamics of thin isolated disks (without counterrotation) have shown that if the disk is too cold in the vertical direction (with $\czcr < 0.3$) it may be subject to a kind of {\it fire-hose} instability. Early investigations were based on a linear theory~(\citealt{1966GFD...66..111T,1977SvAL....3..134P}; see also \citealt{1971Ap&SS..14...52K}). This has been confirmed by many later articles, by means of analytical investigations and by N-body simulations that have also explored the conditions of instability in less idealized models~\citep{1994ApJ...425..551M,2005AstL...31...15S,2010AN....331..731K,2011MNRAS.415.1259G,2013MNRAS.434.2373R} and extended our understanding to the nonlinear evolution of such systems. In general, it has been shown that the excited bending waves (they can be axisymmetric or non-axisymmetric) tend to heat up and to thicken the disk, so as to remove the source of instability; in some cases the $\czcr$ ratio has been observed to increase to values $\approx 0.7-0.9$. Note that fire-hose related bending of stellar disks~\citep{1996ApJ...473..733S} has also been associated with nonlinear processes during bar growth~\citep{1991Natur.352..411R}.

We may state that so far, in relation to the problem of disk heating and thickening, the attention has been drawn mostly to collective phenomena associated with fire-hose instabilities or spiral density waves, and the possible role of counterrotation has been either ignored or overlooked. The main goal of this paper is to find the connection between the presence of counterrotation and the vertical structure of the stellar disk. By means of high-resolution collisionless $N$-body simulations, we investigate the evolution of the vertical-to-radial velocity dispersion ratio $\czcr$ in a set of models with different masses and initial conditions for the pressure tensor.

The paper is organized as follows. In Sect.~\ref{sec::model} we describe the models considered in our numerical simulations, especially in relation to a basic set of parameters that are expected to characterize their stability properties. In  Sect.~\ref{sec::results} we present and discuss the results of our numerical simulations. We start by considering cases without counterrotation, so as to exclude the interference of Jeans instability effects from the following runs with counterrotation~(Subsect.~\ref{sec::results_corot}). The general evolution of models with counterrotation is described in detail in Subsect.~\ref{sec::results_counter}. The particular behavior related to the evolution of the $\czcr$ ratio is presented in Subsect.~\ref{sec::results_general}. In the last two sections, Sects.~\ref{sec::discuss} and~\ref{sec::concl}, we address some closely related issues and summarize the results of the paper.

\section{Models and parameter regimes}\label{sec::model}

\subsection{Setting up the simulations}\label{secc::code}

We consider the full three-dimensional dynamics of two-component stellar disks embedded in a fixed gravitational potential representing an inactive, axisymmetric dark matter halo. We used the $N$-body code described in~\cite{2014JPhCS.510a2011K} where gravity force in the disk is calculated by the TreeCode Top Down algorithm~\citep{1986Natur.324..446B}. The adopted scheme of the fourth order provides a sufficiently accurate integration of the equations of the motion. The code was
extensively used to resolve the issues of spiral structures formation in pure collisionless systems~\citep{2012MNRAS.427.1983K,2013MNRAS.431.1230K} and in a combined $N$-body/hydrodynamical simulation of galaxy evolution~\citep{2012ARep...56..664K,2016MNRAS.455.1782K}.  In the present work the opening angle parameter of the TreeCode algorithm is assumed to be $0.1$. In physical units, the integration step is chosen to be $ 2 \times 10^4$~yr. We consider simulation particles of the same mass (in the range $10^3-10^4$~\Msun) for both components; the total number of particles is in the range $2-3\times 10^6$. 

The initialization starts from a specification of the projected density distributions $\Sigma_1$ and $\Sigma_2$, taken to be exponentially declining with $R$, with exponential scale lengths $h_1$ and $h_2$. The relevant gravitational forces, including the contribution of the fixed halo potential, are then computed by means of the tree-code. At this stage we determine circular velocities and then proceed to assign the radial velocity dispersions. For each component the radial velocity dispersion $\sigma_R (R)$ is taken to have an exponential profile with scale length twice that of the corresponding projected density profile. As a result, for a single isolated disk and a choice of vertical velocity dispersion $\sigma_z \propto \sigma_R$, this would naturally lead to a vertical density profile of the $\cosh^{-2}{\left( z/z_0\right)}$ form, characterized by an approximately constant thickness $z_0$, with $z_0 = \sigma^2_z / (\pi G \Sigma)$. Obviously, when two different disk components and an external halo coexist, the relations mentioned above apply only approximately (in particular, for each component it is possible to represent the vertical density profile with an approximate $\cosh^{-2}$ profile in which the projected density determining the related thickness $z_{0i}$ is the total density $\Sigma_{\rm tot} = \Sigma_1 + \Sigma_2$). In any case, starting from the approximate conditions listed above, the vertical equilibrium is generated with the help of a few iterations of the Jeans equations for the two components~\citep{2001ARep...45..180K}. We are thus able to consider models that are initially characterized by various central surface densities, various exponential scale lengths, and various initial $\czcr$ ratios for the two disk components.

There is no evidence yet of bars and
strong spirals in galaxies with counterrotation. Typically both components of galaxies with counterrotation look like lenticular galaxies~\citep{2014ASPC..486...51C}. Observations suggest that the secondary (less bright) component is characterized by a line-of-sight velocity dispersion lower  than that of the host component~\citep[e.g., see][]{2016MNRAS.461.2068K}. These empirical facts are taken into consideration in our numerical models that we thus hope are sufficiently realistic. Therefore, to isolate the effects of counterrotation on the vertical disk heating, we investigate models where bar instability is suppressed.

\subsection{Relative properties of the two disks}\label{secc::rel}

To characterize the relative properties of the two components, we follow the notation introduced by~\citet{2006A&A...459..333B} (in the context of a zero-thickness two-fluid model). In particular, 
we define the following dimensionless parameters, in which the index $1$ corresponds to the more massive host component and $2$ corresponds to the secondary component:
\begin{itemize}
\item surface density ratio
\begin{equation}
\alpha \equiv \Sigma_2 / \Sigma_1\,;\label{eq::alpha}
\end{equation}
\item radial velocity dispersion (``temperature") ratio
\begin{equation}
\beta \equiv \sigma^2_{\rm R2}/\sigma^2_{\rm R1} \,;\label{eq::beta}
\end{equation}
in general, we consider $\beta<1$, that is, we assume that the secondary component is colder.
\item angular velocity ratio 
\begin{equation}
\delta \equiv \Omega_2/\Omega_1 \,,\label{eq::delta}
\end{equation}
where $\delta>0$ denotes corotating disks and $\delta<0$ characterizes the case in which the secondary component is counterrotating. 
\end{itemize}

\subsection{Some average properties of the two-component system}\label{secc::heat}

To quantify the local and global heating of the two-component system in the course of the simulations we proceed in the following way. By using radial profiles of the velocity dispersion and disk thickness we can calculate one-component radially-averaged quantities:
\begin{equation}
\langle X_{1,2} \rangle = \int\limits^\infty_0 2\pi r X_{1,2}(r) \Sigma_{1,2}(r)dr / \int\limits^\infty_0 2\pi r \Sigma_{1,2}(r)dr   \,,\label{eq::X}\
\end{equation} 
where $X$ is the radial velocity dispersion $\sigma_{\rm R}$, vertical velocity dispersion $\sigma_{\rm z}$, vertical to radial velocity dispersion ratio $\czcr$, or vertical disk thickness $z$; index 1 corresponds to the host component, index 2 to the secondary component. 

We also introduce two-component averaged quantities:
\begin{equation}
\Oo  \left\langle Y \right\rangle   = M^{-1}_{\rm tot}  \int\limits^\infty_0 2\pi r \left[ Y_1(r) \Sigma_1(r) + Y_2 (r) \Sigma_2(r) \right] dr\,.\label{eq::Y}
 \end{equation} 
where $Y$ is the radial velocity dispersion $\sigma_{\rm R}$, vertical velocity dispersion $\sigma_{\rm z}$,  vertical disk thickness $z$, or vertical to radial velocity dispersion ratio $\czcr$, and the total disk mass is:
\begin{equation}
M_{\rm tot} =  \int\limits^\infty_0  2\pi r [ \Sigma_1(r)  + \Sigma_2(r)] dr.\label{eq::total_mass}
\end{equation}
As already noted, in the course of the simulations, when the system develops non-axisymmetric features, the above definitions are meant to include an average with respect to the azimuthal direction~$\varphi$. 

\subsection{Parameters relevant to the development of density waves}\label{secc::dw}

With respect to axisymmetric density waves in a one-component disk, it is well known~\citep[][]{1964ApJ...139.1217T} that fluid and stellar disks are characterized by a similar definition of the stability parameter $Q$, provided that the dispersion $\sigma$ that in the fluid model enters through the definition $Q = \sigma \kappa/ (\pi G \Sigma)$ be replaced, in the stellar disk, by $\approx \sigma_{\rm R} (\pi/3.36)$. 

It is also well known that for multi-component systems the corresponding stability criterion is more complicated. Several studies have addressed the issue for various cases, given the fact that each component can be gaseous or stellar~\citep[][and references therein]{1966PNAS...55..229L,1996ssgd.book.....B,2011MNRAS.416.1191R,2013MNRAS.433.1389R,2014dyga.book.....B}.
For a zero-thickness fluid model of a two-component disk (without counterrotation) the marginal stability curves in the $(\hat{\lambda}, Q^2_1)$ plane, where $\hat{\lambda}$ is the relevant dimensionless radial wavelength of the perturbation, have been calculated for different values of the parameters 
$(\alpha, \beta)$ that define the density and temperature ratios of the two disks~\citep[see Eq.~(16.9) in][]{2014dyga.book.....B}. By introducing the effective stability parameter:
\begin{equation}
\Qeff = \frac{Q_1}{Q_{\rm max}(\alpha,\beta)}\,,\label{eq::Qeff}
\end{equation}
where $Q_{\rm max}(\alpha,\beta)$ is the maximum value of the marginal stability curve, the threshold of axisymmetric Jeans instability then occurs at 
$\Qeff = 1$. In general, $Q_{\rm max} > 1$,  because the addition of a secondary component is destabilizing. For dynamically coupled components the marginal stability curve has a single peak and its shape remains similar to that of one-component disks. On the other hand, even a light secondary component, if sufficiently cool, can have a significant impact on the stability of the combined two-component disk. In other words, each component can apparently be well on the stable side, with $Q_{1,2} > 1$, yet the combined disk can be gravitationally unstable with respect to axisymmetric Jeans instability, with $\Qeff < 1$. This general statement has been proved for both stellar-fluid models and fluid-fluid models. Under certain circumstances, especially when the secondary component is too cold, the two components may become dynamically decoupled, which is marked by the fact that the marginal stability curve changes shape, into a curve with two peaks.

On the other hand, with respect to non-axisymmetric density-wave perturbations~\citep[][and references therein]{1981seng.proc..111T, 2014dyga.book.....B}, stability depends on a second parameter, which describes the effects of a new instability mechanism related to shear (the mechanism is generally called swing or overreflection, and, for a given value of the azimuthal wavenumber $m$, it starts to operate when the disk is sufficiently heavy; e.g., see Eq.~(15.33) in \citet{2014dyga.book.....B}). Therefore, density waves can be locally unstable even for relatively large values of $Q$ (e.g., see Fig.~15.9 and Eq.~(15.34) in \citet{2014dyga.book.....B}). In simulations with corotating components we then expect that, even if initialized in such a way that $\Qeff > 1$, some heating should occur in the plane as a result of spiral activity. In a collisionless disk this heating process may be partly limited by the presence of Lindblad resonances; however, this behavior is difficult to reproduce by means of $N$-body simulations, because simulations tend to lack the desired resolution in phase space. In contrast, in simulations with counterrotating disks, even if by adding a secondary component the mass of the disk is increased, the heating related to spiral activity due to overreflection may be less important, because the overreflection of non-axisymmetric waves, which relies on the presence of coherent shear, is likely to be jeopardized by the fact that the added secondary component rotates in the opposite direction with respect to the primary component. 

\begin{figure}[h!]
\includegraphics[width=1\hsize]{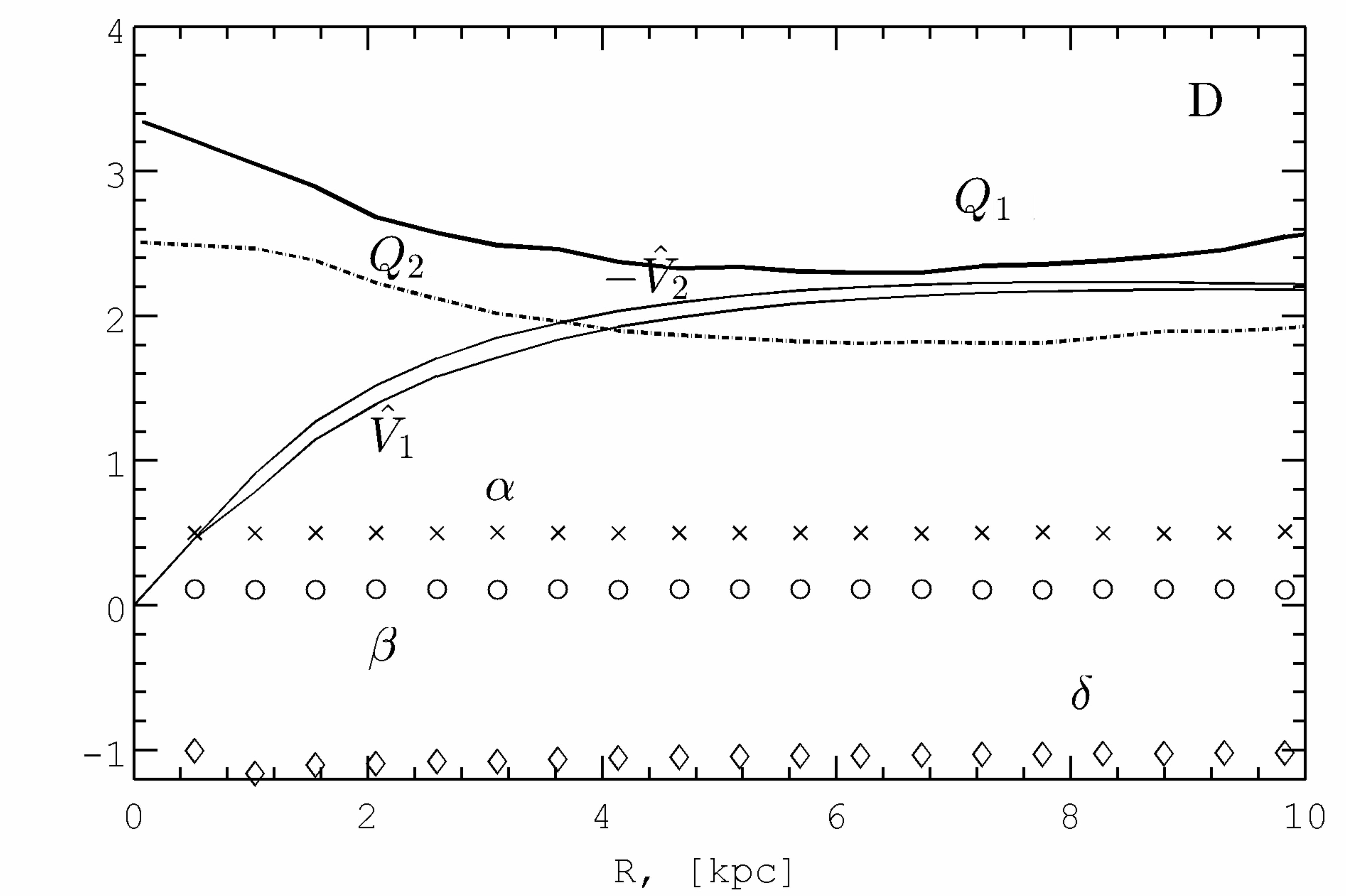}
\caption{Radial profiles of the initial parameters for model D (with counterrotation). The surface density ratio $\alpha$, the radial temperature ratio $\beta$, the angular velocity ratio $\delta$, and the axisymmetric stability parameters $Q_{\rm 1}(R)$ and $Q_{\rm 2}(R)$ associated with the two components. The curves $\hat{V}_{1,2}$ are the rotation curves of the two components, in units of 100~\kmps}.\label{fig::ini}
\end{figure}

In Fig.~\ref{fig::ini} we illustrate the radial profiles of some of the quantities defined above, for model D (characterized by counterrotation). In Table~\ref{tab::tabular1} we list some properties of the initial models considered in the simulations that we have performed; the last three columns provide values of the listed parameters taken at the radial location $R = R^*$ where $Q_1$ has its minimum.

\begin{table*}
\caption{Initial parameters of different runs. Models without counterrotation are marked by $\dagger$ ($\delta=1$). Here $\Sigma_1(R=0)$ is the central value of the surface density of the primary component, $h_2$ is the exponential scale length of the density distribution of the second component (for the first component, $h_1=3$~kpc), $\sigma_{\rm R1}(R=0)$ and $\sigma_{\rm R2}(R=0)$ are the central values of the velocity dispersion for each component, $Q_{\rm 1}(R = R^*)$ is the minimum value of the one-component stability parameter of the host component, and $\Qeff(R = R^*)$ is the effective two-component stability parameter according to Eq.~(\ref{eq::Qeff}) evaluated at $R^*$. Initial values of the velocity dispersion ratio for the two components are constant with radius for all models. The definitions of the density ratio $\alpha$ and of the temperature ratio $\beta$ are given in the text; also these parameters are generally constant with radius, except for models Kb and Kc, for which the values listed here are taken at $R = R^*$, that is, the location where $Q_1$ has its minimum. Model I10 considers the case of exactly equal but counterrotating disks, which is unlikely to occur in real systems.}
\begin{center}
\begin{tabular}{lcccccccccccccccc}
\hline
 \\
Model & $\Sigma_1(R=0) $ & $h_2$ & $\sigma_{\rm R1}(R=0)$ & $\sigma_{\rm R2}(R=0)$ & $\Oo\frac{\sigma_{\rm z1}}{\sigma_{\rm R1}}$ & $\Oo\frac{\sigma_{\rm z2}}{\sigma_{\rm R2}}$& $\alpha$ & $\beta$  & $Q_{\rm max}(\alpha,\beta)$ & $Q_1(R = R^*)$ &  $\Qeff(R = R^*)$ &  \\
 & \Msunpc  & kpc & \kmps & \kmps &\\
\hline
A$^\dagger$  & 750  & 3 & 60 & 35 & 0.5 & 0.5 &0.5 & 0.34  & 1.8 & 1.44  & 0.8 \\
B$^\dagger$  & 750  & 3 & 100 & 35& 0.5 & 0.5 & 0.5 & 0.12  & 2.2 & 2.2 &  1.0 \\
\hline
C  & 750 &  3 & 60 & 35& 0.5 & 0.5 & 0.5 & 0.34  & 1.8 &  1.44 &  0.8 \\
D  & 750 & 3 & 100 & 35& 0.5 & 0.5 & 0.5 & 0.12  & 2.2 & 2.2 & 1.0 \\
E  & 400 &  3 & 100 & 35& 0.5 & 0.5 & 0.5 & 0.12  & 2.2 & 3.96  & 1.8 \\
G  & 400 &  3 & 60 & 35& 0.5 & 0.5 & 0.5 & 0.34  & 1.8   & 2.8 &  1.5 \\
\hline
I01  & 1050 &  3 & 50 & 50& 0.5 & 0.5 & 0.1 & 1  & 1.08 & 1.0 &   1.0 \\
I02  & 970 & 3 & 50 & 50& 0.5 & 0.5  & 0.2 & 1  & 1.2 & 1.2 &   1.0 \\
I03  & 895 & 3 & 50 & 50& 0.5 & 0.5 & 0.3 & 1  & 1.3 & 1.3 &   1.0 \\
I04  & 830 & 3 & 50 & 50& 0.5 & 0.5 & 0.4 & 1  & 1.4 & 1.4 &   1.0 \\
I05  & 775 & 3 & 50 & 50& 0.5 & 0.5 & 0.5 & 1  & 1.5 & 1.5  & 1.0 \\
I07  & 660 & 3 & 50 & 50& 0.5 & 0.5 & 0.7 & 1  & 1.7 & 1.7 &   1.0 \\
I10  & 525 & 3 & 50 & 50& 0.5 & 0.5 & 1 & 1  & 2.0 & 2.7 &   1.4 \\
\hline
Kb  & 750  & 2 & 90 & 30& 0.7 & 0.4 & 0.5 & 0.12  & 1.48 & 1.48   & 1.0 \\
Kc  & 750 & 5 & 90 & 30& 0.6 & 0.2 & 0.5 & 0.1  & 1.51 & 1.51 &   1.0 \\
\hline
\end{tabular}\label{tab::tabular1}
\end{center}
\end{table*}

In conclusion, for the majority of models we set up the initial parameters in such a way that $\Qeff(R) > 1$. We also study a case in which, during evolution, Jeans instability should coexist with the effects due to counterrotation (model C).

As to the linear stability analysis for tightly-wound non-axisymmetric perturbations in the plane of the disk related to the two-stream instability, we may refer to the study carried out by \cite{2006A&A...459..333B}~who examined the dispersion relation of a zero-thickness two-fluid model in the WKB  (tightly wound) approximation (see Eqs.~(4)-(7) in that article); the general formulation makes the analysis applicable to both cases of corotating and counterrotating components. 
Note that the analysis demonstrates that axisymmetric (i.e., $m = 0$) perturbations do not distinguish the cases of corotating and counterrotating components, with the stability condition given by~(Eq.~\ref{eq::Qeff}). To compare the stability of a disk made of corotating components with that of a disk made of counterrotating components, we set $m \neq 0$ (and $\eta \approx 0.01$ for models with corotation and $\eta \approx -1.4$ for models with counterrotation), and consider the solutions of the dispersion relation~\citep[Eq.~(4) in][]{2006A&A...459..333B}. For the definition of $\eta$, see the article just cited.

\begin{figure*}
\includegraphics[width=0.5\hsize]{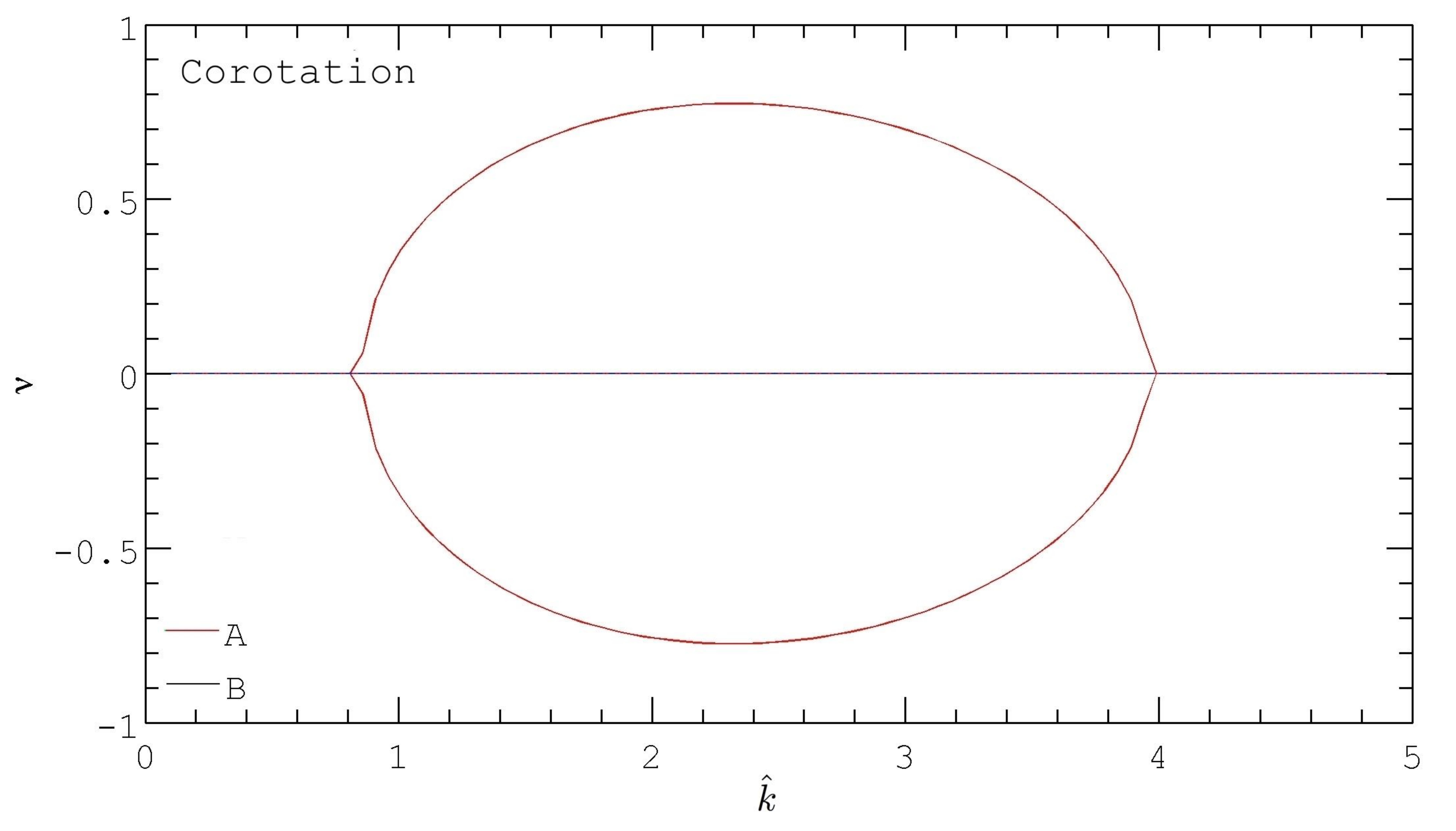}\includegraphics[width=0.5\hsize]{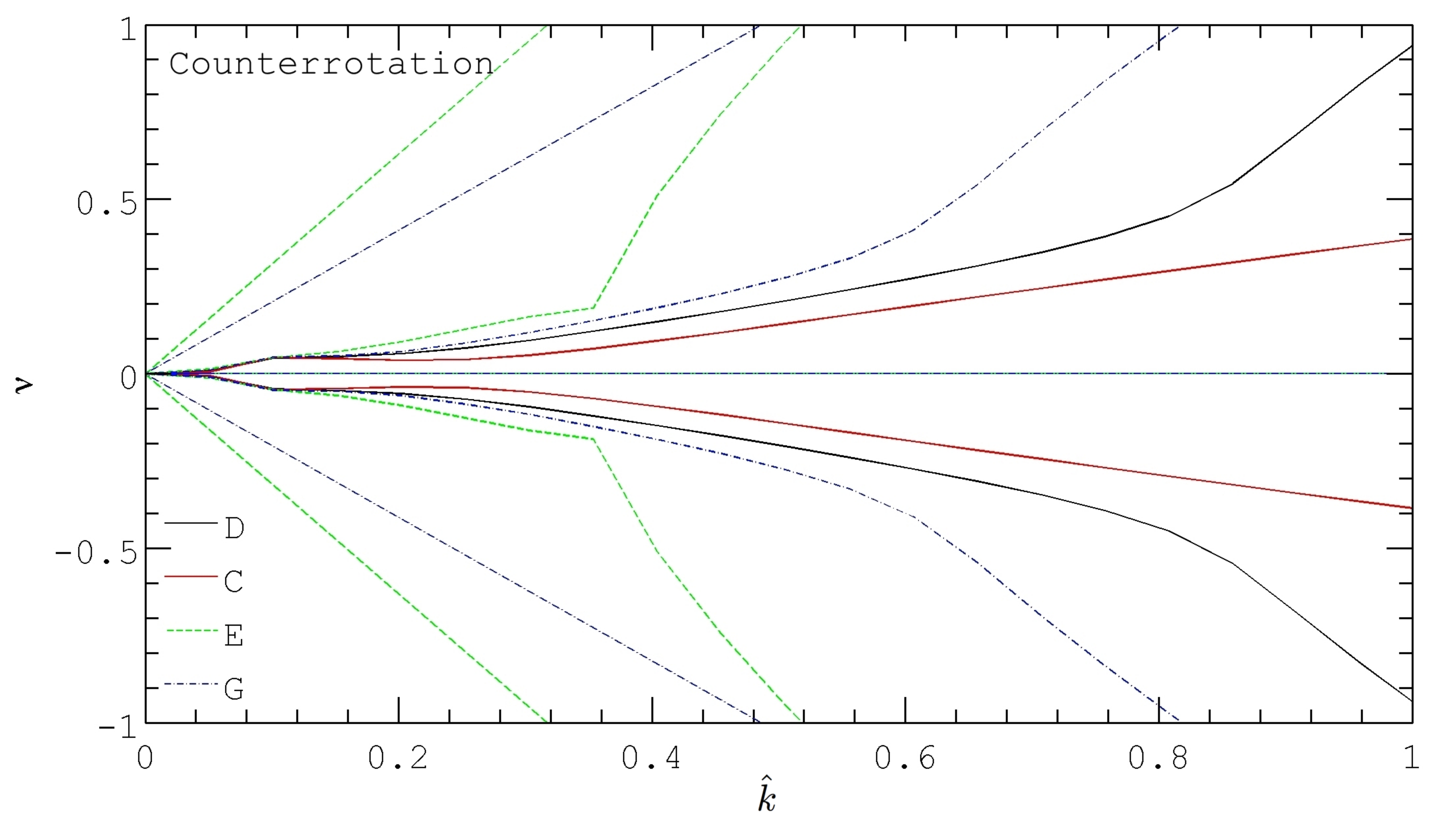} \caption{Imaginary part of the solutions of the dispersion relation for two-component disks with 
$\eta \approx 0.01$ for models without counterrotation~(left frame) and $\eta \approx -1.4$ with counterrotation (right frame), where $\hat{k}$ and $\nu$ are the relevant dimensionless radial wavenumber and the dimensionless Doppler-shifted frequency, respectively (following the notation of  \cite{2006A&A...459..333B}).}\label{fig::imag_disp_rel}
\end{figure*}

In Fig.~\ref{fig::imag_disp_rel} we plot the imaginary parts of the solutions of the dispersion relation for pairs of models differing only by the direction of rotation of the secondary component. In the corotating case, only one model has a prominent unstable root (model A, $\Qeff(R=R^*)<1$), which we interpret to be the result of Jeans instability. Counterrotation makes all the models locally unstable. The linear dispersion relation of each model with counterrotation has several unstable roots, dependent on the value of $m$ (or $\eta$). The nonlinear evolution induced by all these instabilities can only be followed by means of simulations. 

\subsection{Issues related to the stability of bending waves}\label{secc::bending_stab}

As to the properties of bending waves, linear stability analyses focusing on the role of the free energy associated with the relative motion between two components have been carried out in many papers \citep[e.g., see][]{1971ApL.....9...37M,	1980A&A....88..289B, 1982A&A...106..274B, 1987AJ.....94...99A}. As with the discussion of the stability given in the previous subsection, beyond the simple realization that counterrotation has a destabilizing role, the main difficulty is to make quantitative predictions on the resulting nonlinear evolution. Therefore, we will not discuss further the related issues; instead, we decide to proceed directly to the results of our numerical simulations.

\section{Results of simulations}\label{sec::results}
\subsection{Preliminary simulations without counterrotation}\label{sec::results_corot}
Because we wish to identify effects clearly induced by the presence of counterrotation, we first consider two-component corotating disks (models A and B in Table~\ref{tab::tabular1}) to find conditions under which the disk is reasonably free from major instabilities.

\begin{figure*}
\includegraphics[width=1\hsize]{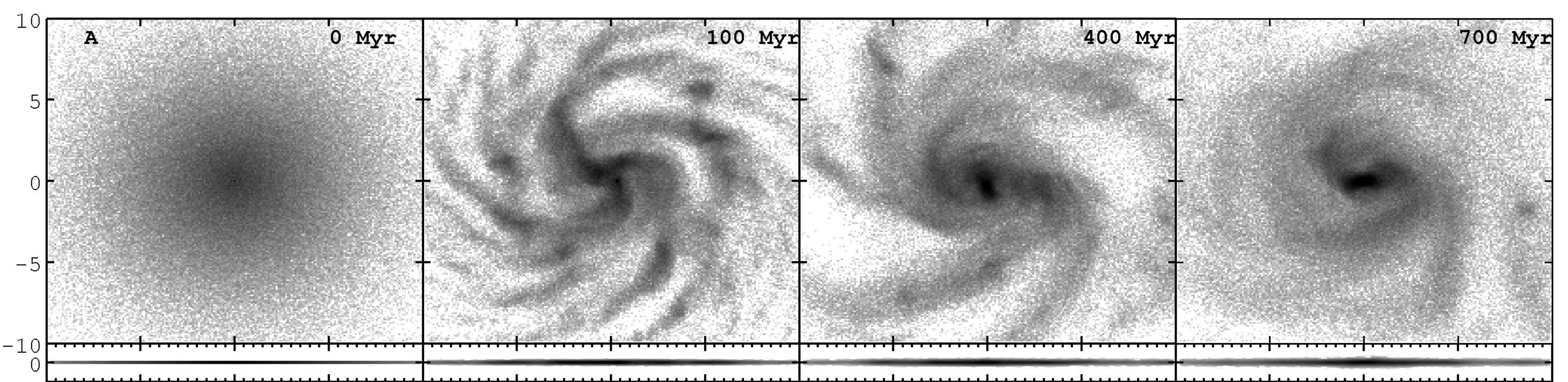}
\includegraphics[width=1\hsize]{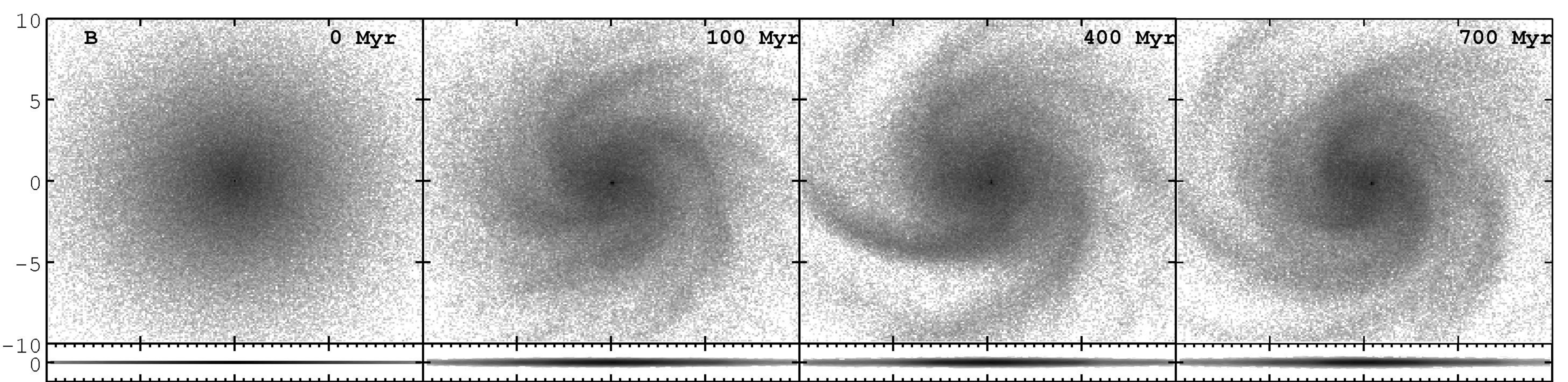} 
\caption{Evolution of two-component corotating disks: top row illustrates the behavior of model A ($\Qeff(R=R^*)=0.8$), bottom row describes the evolution of model B ($\Qeff(R=R^*)=1$). Edge-on and face-on maps are shown on the same spatial scale. Face-on images have 20~kpc spatial size, vertical size of edge-on views is equal to 4~kpc.}\label{fig::corot_evol}
\end{figure*}

The effects of possible axisymmetric instabilities in the disk plane is under control, as can be seen from Fig.~\ref{fig::corot_evol}, where we illustrate the evolution of model A (expected to be locally unstable with respect to axisymmetric perturbations, $\Qeff(R=R^*)=0.8$) and model B (expected to be locally stable with respect to axisymmetric perturbations, $\Qeff(R=R^*)=1.0$). The values of $\Qeff(R = R^*)$ should be taken as indicative only, because the definition of this effective stability parameter is taken from a zero-thickness two-fluid model, whereas the simulations that we perform aim at describing the three-dimensional behavior of two-component stellar disks. Because of physical mechanisms that can be traced to overreflection (or swing; see discussion in Subsect.~\ref{secc::dw}), multiple-armed spiral activity is present in both models. Figure~\ref{fig::corot_evol} also shows the absence of bending waves. In a few rotation periods the disk heats up and evolves until phenomena ``saturate"~\citep[see details in][]{1997ApJ...477..410L,2012MNRAS.427.1983K}. The disk thickens during evolution, but its vertical structure does not change much, especially in model B.

\begin{figure*}
\includegraphics[width=0.5\hsize]{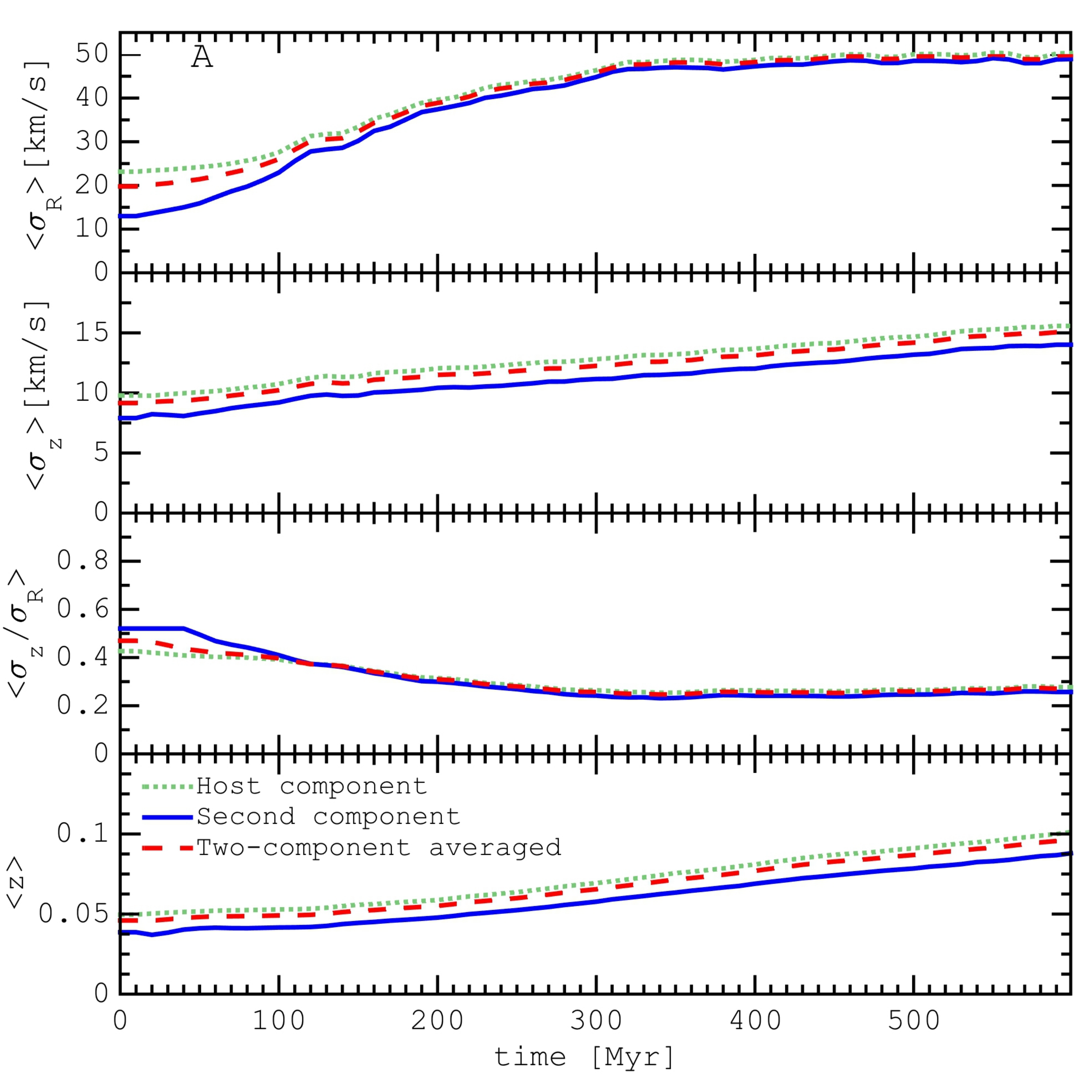}
\includegraphics[width=0.5\hsize]{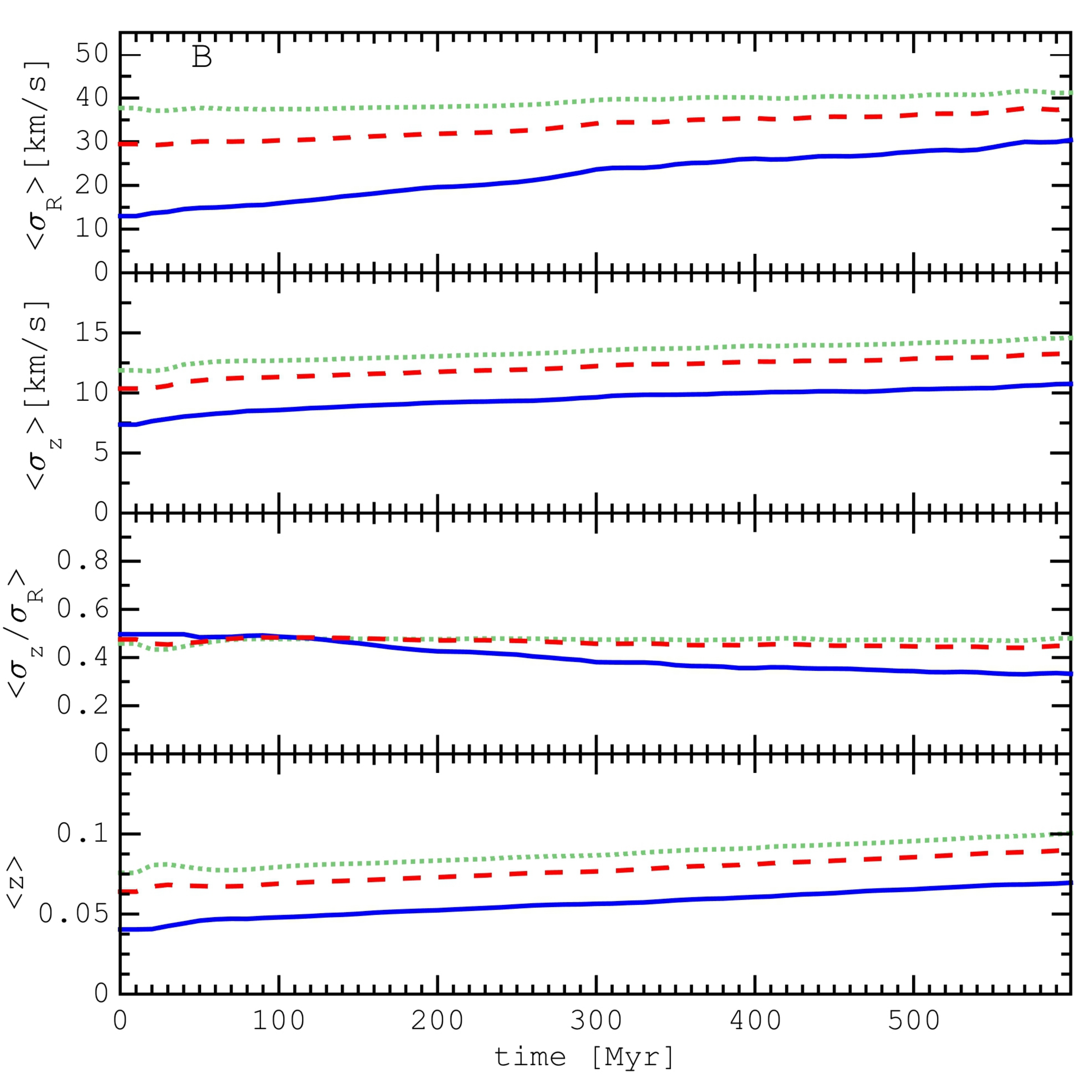} 
\caption{ Evolution of the disk parameters: radial velocity dispersion $\langle \sigma_{\rm R1}\rangle$ , $\langle\sigma_{\rm R2}\rangle$, $\langle\sigma_{\rm R}\rangle$ (first row), vertical velocity dispersion $\langle\sigma_{\rm z1}\rangle$, $\langle\sigma_{\rm z2}\rangle$, $\langle\sigma_{\rm z}\rangle$~(second row), vertical-to-radial velocity dispersion ratio $\czcrmo$, $\langle \sigma_{\rm z2}/\sigma_{\rm R2}\rangle$, $\czcrm$~(third row) and disk thickness $\langle z_1\rangle$, $\langle z_2\rangle$, $\langle z\rangle$ (bottom row); see Eqs.~(\ref{eq::X},\ref{eq::Y}). Blue circles correspond to the
secondary (less massive) component, green crosses to the host component, and red squares represent the two-component averaged values. We recall that the two components in these models are corotating. }\label{fig::corot_evol_params}
\end{figure*}

In Fig.~\ref{fig::corot_evol_params} we show that for model A the radially-averaged values of the radial velocity dispersion~($\langle\sigma_{\rm R1}\rangle$, $\langle\sigma_{\rm R2}\rangle$, and the two-component average $\langle\sigma_R\rangle$), increases from 10-20 up to 50~\kmps. The radially-averaged disk thickness and velocity dispersion ratio do not change significantly during the evolution even in the unstable model A. This may be due to the relatively large initial velocity dispersion ratio (0.5 for both components), well above the critical value $0.3-0.37$ for~(fire-hose) bending instabilities~\citep{2000AstL...26..277R}. Note that after several dynamical times the vertical-to-radial velocity dispersion ratio decreases as a result of heating in the plane associated with spiral density-wave activity. Such effect is well known and has been demonstrated numerically, for example, by ~\cite{2005ApJ...629..797G,2006ApJ...645..209D,2012MNRAS.427.1983K}. Other two-component corotating disks~(models F and G) are even more stable, because their effective stability parameter is larger $\Qeff(R=R^*) \gtsima 1.5$ (actually, for sufficiently large values of $\Qeff(R=R^*)$ we expect that even multi-armed spiral activity should be absent). In principle, the two-component corotating disks might be slightly unstable, because of the possible influence of the subtle mechanism associated with the asymmetric drift explored by \cite{2006A&A...459..333B}; however, a quantitative analysis of such effects is beyond the scope of the present work. 

We conclude that, when counterrotation is kept ``turned-off", our models are generally free from the effects of bending instabilities and strong vertical heating. Therefore, in the following subsection we can focus on effects that we can safely attribute to the presence of counterrotation.

\subsection{Simulations with counterrotation}\label{sec::results_counter}
We start by considering two-component models with disks characterized by the same parameters as those of the models described in the previous section, but in the case when the two disks are in mutual counterrotation. We pay special attention to the vertical structure of the disk. 

\begin{figure*}
\includegraphics[width=1\hsize]{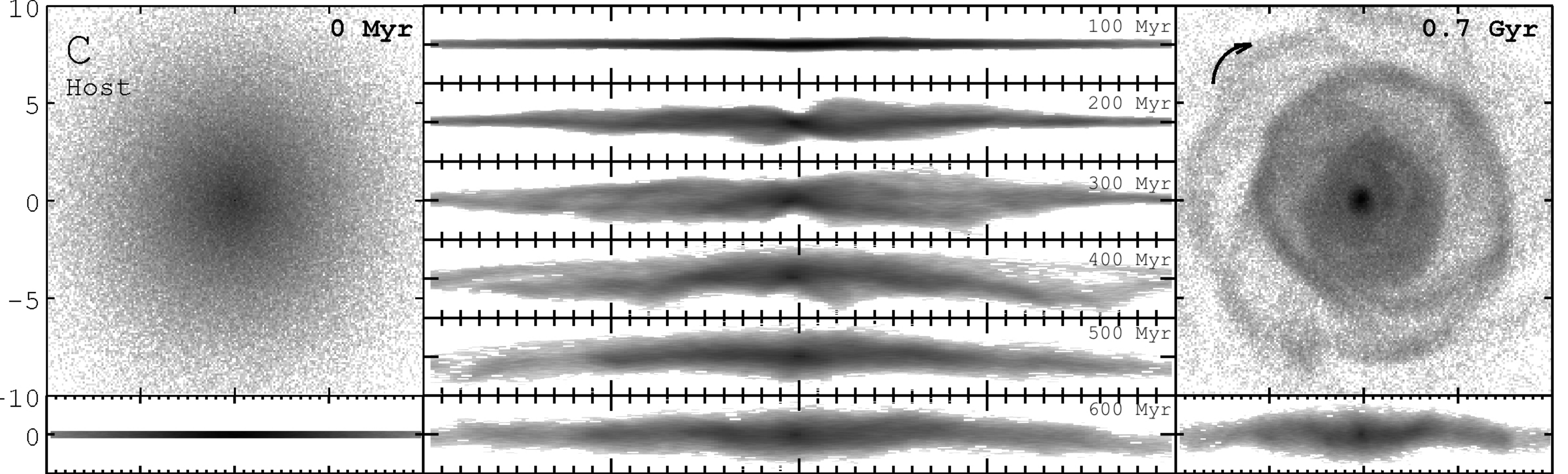}
\includegraphics[width=1\hsize]{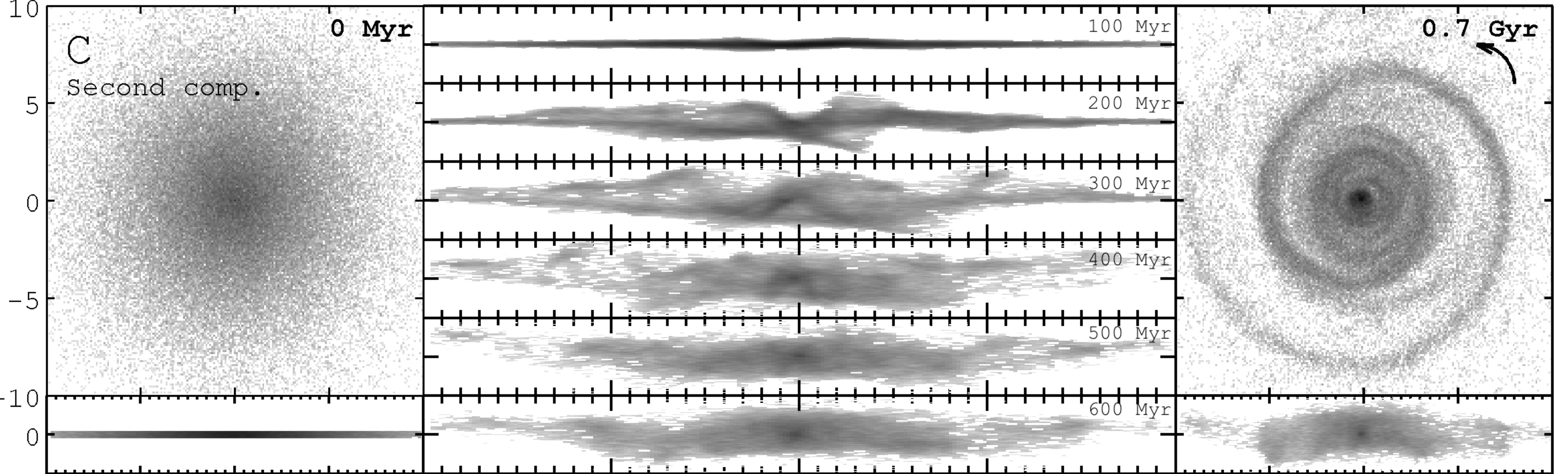}
\includegraphics[width=1\hsize]{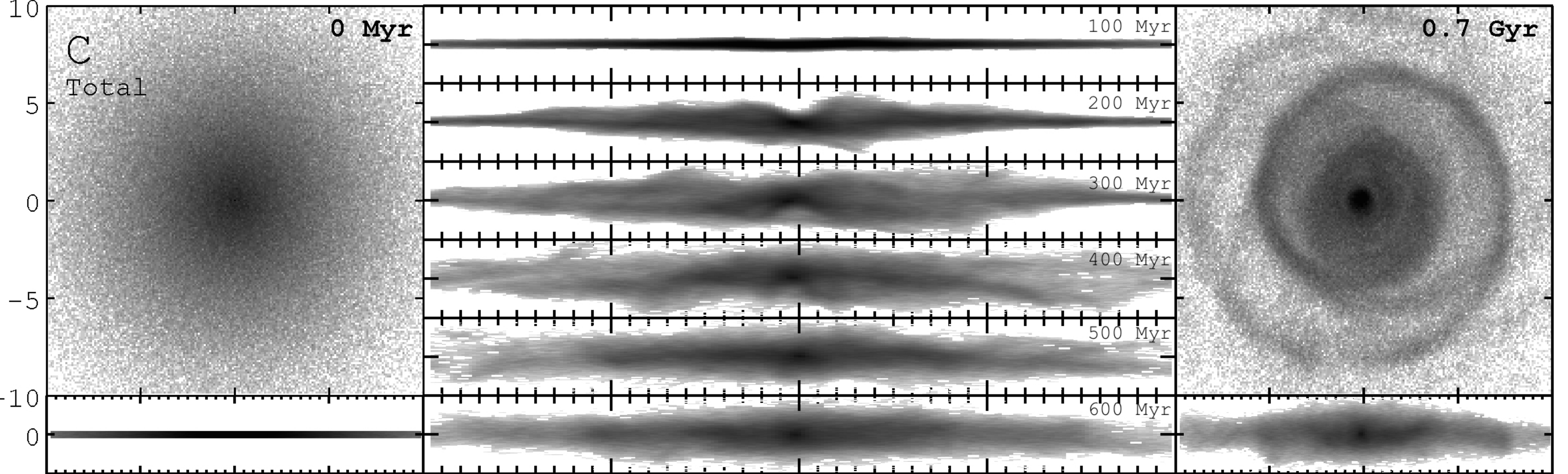}
\caption{Evolution of the unstable ($\Qeff(R=R^*)=0.8$) model C in the presence of counterrotation. In this composite figure, the top frames 
illustrate the evolution of the host component, the middle frames the evolution of the secondary counterrotating component, and the bottom frames display the appearance of the disk obtained by superposition of the two components. The left part of the figure shows an initial snapshot of the system, whereas the right part illustrates the face-on morphology of the system at time $t = 0.7$ Gyr. The central part of the figure illustrates the time-dependent evolution of the vertical disk structure (edge-on view) in the time interval $100-600$ Myr, with steps of 100 Myr. Face-on images have 20~kpc spatial size, vertical size of edge-on views is equal to 4~kpc.}\label{fig::conterrot_evol1}
\end{figure*}

\begin{figure*}
\includegraphics[width=1\hsize]{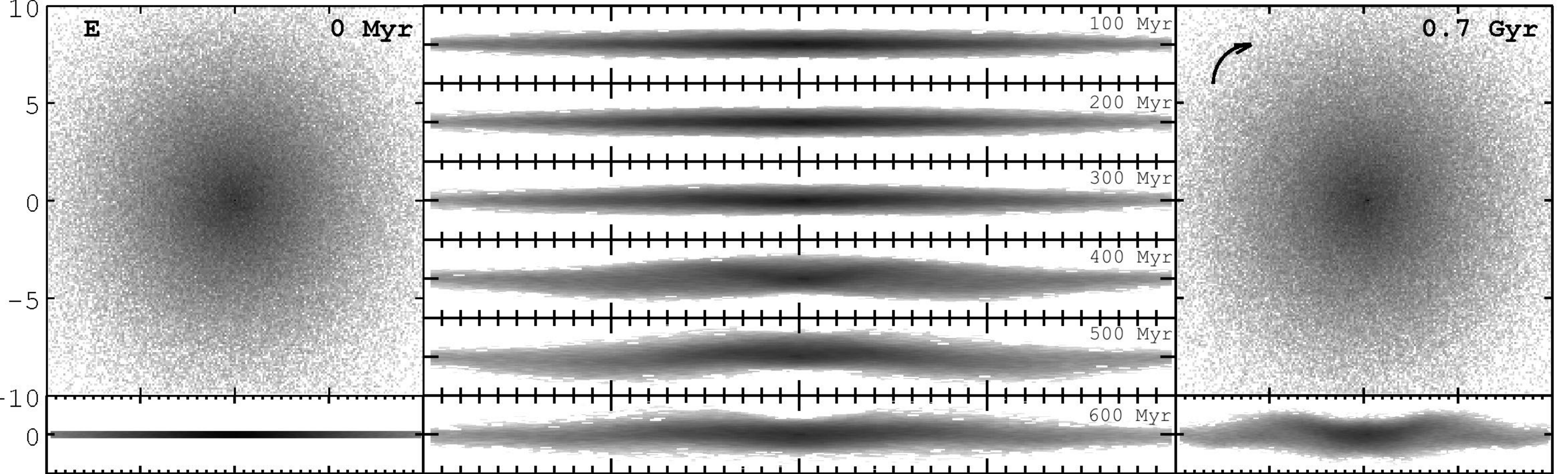}
\includegraphics[width=1\hsize]{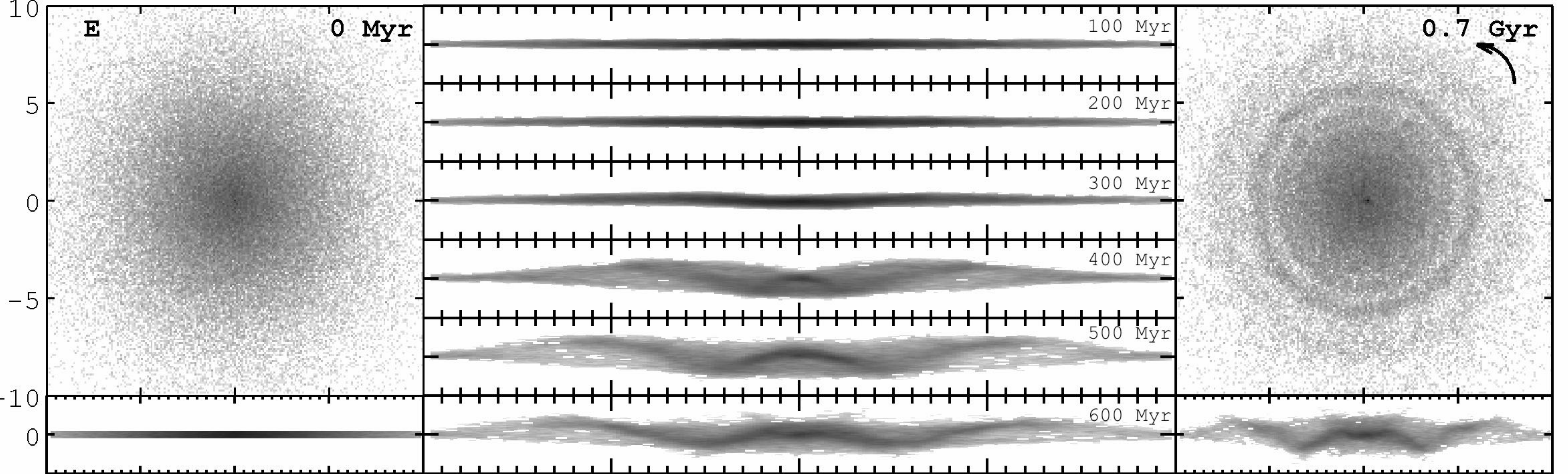}
\includegraphics[width=1\hsize]{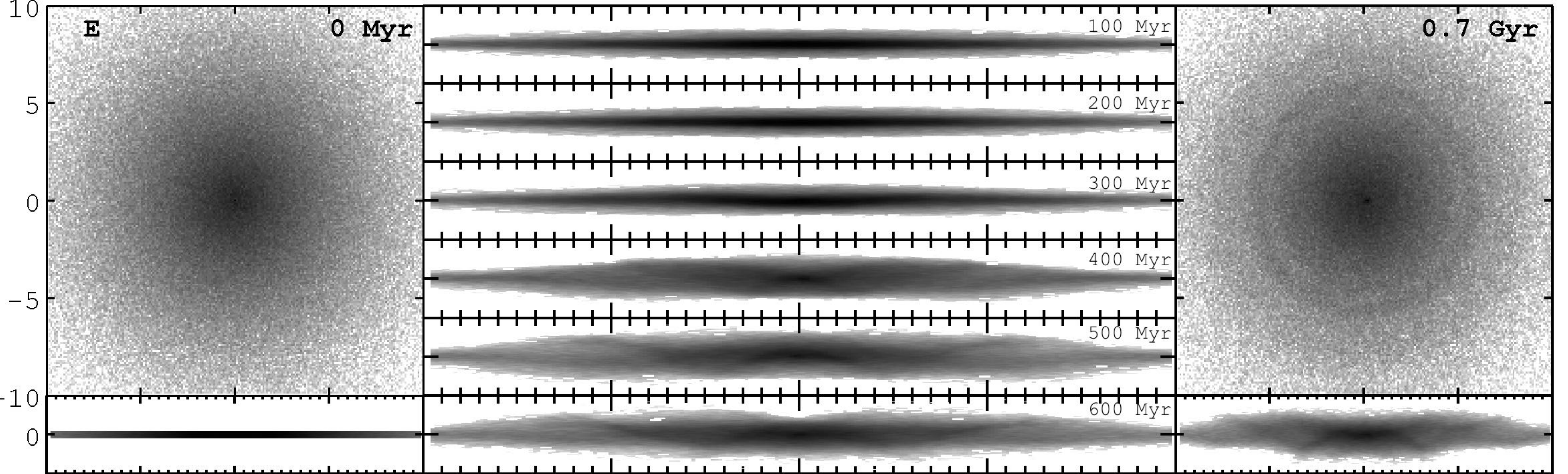}
\caption{The same as in Fig.~\ref{fig::conterrot_evol1}, but for model E ($\Qeff(R=R^*)=1.8$).}\label{fig::conterrot_evol2}
\end{figure*}

Figures~\ref{fig::conterrot_evol1}, \ref{fig::conterrot_evol2} show the evolution of two very different models. Model C is expected to be very unstable, because counterrotation operates in the presence of a Jeans-unstable situation ($\Qeff(R=R^*)=0.8$); in turn, model E is characterized by $\Qeff(R=R^*)=1.8$. In both cases we observe the excitation of bending waves as well as the formation of various structures in the face-on view of the plane.

The more unstable model exhibits the formation of a coherent one-armed spiral pattern (see Fig.~\ref{fig::conterrot_evol1}). If referred to the rotation axis of the primary component, the observed structure is {\it leading}, which agrees with the results of the 2D simulations by~\cite{1997ApJ...484L..33C}. This is likely to occur as a result of the combined effect of gravitational instability and counterrotation, because the phenomenon is not observed in model E (see Fig.~\ref{fig::conterrot_evol2}), for which only symmetric rings are clearly seen in the face-on view of the disk. In the face-on view of model C, in addition to the dominant one-armed structure several small-scale patterns are present in the primary component. For model C, the excitation of the bending waves starts from the center~(see Fig.~\ref{fig::conterrot_evol1}).At 200 Myr bending waves are already seen clearly in the center, where the initially thin disk rapidly becomes thicker. Further evolution and thickening affects the outer regions. Bending waves appear to propagate with opposite phases relative to the disk plane in the two components. Note that bending structures are very asymmetric in both components of the disk.
 
For the gravitationaly stable model E~(see Fig.~\ref{fig::conterrot_evol2}), we also see disk thickening and bending wave excitation. However, structures across the disk are much more symmetric than in the previous case. In the face-one view of model E there is only a ring-like structure.

\begin{figure*}
\includegraphics[width=1\hsize]{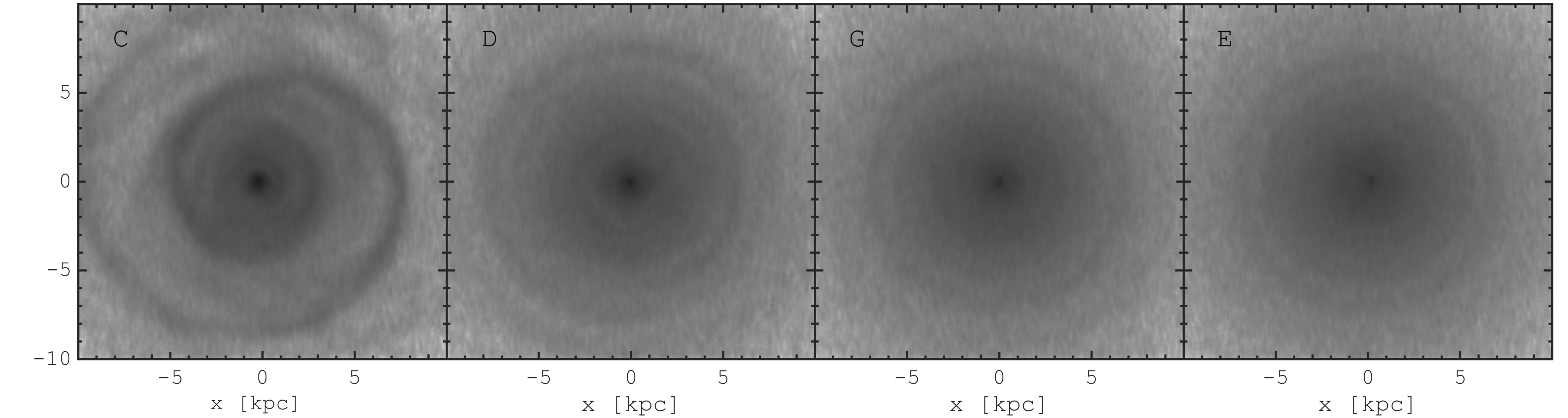}
\caption{ Total stellar surface density distribution in the galactic disk at 700~Myr for the four models with counterrotation.  From left to right: C~($\Qeff(R=R^*) = 0.8$), D~($\Qeff(R=R^*) = 1.0$), G~($\Qeff(R=R^*) = 1.5$), E~($\Qeff(R=R^*) = 1.8$). Each image is 20 kpc across. }\label{fig::rings_faceon}
\end{figure*}

Ring structures are quite surprising, because the linear WKB stability analysis of counterrotation would suggest that $m = 0$ perturbations ``do not see" the presence of counterstreaming. To clarify the origin of the ring-like structures we compare snapshots of the evolution of models with various values of $\Qeff(R=R^*)$. In Fig.\ref{fig::rings_faceon} we show the stellar surface density distributions for models C, D, G, and E. In all models with $\Qeff(R=R^*)\geq 1$ we see the presence of ring-like structures at 5-6~kpc from the center. In models with relatively low $\Qeff(R=R^*)$ (C and D) there are also asymmetric structures, whereas in models with $\Qeff(R=R^*)>1$  pure ring-like structures are seen. Therefore, in our models ring formation is not expected to result from Jeans instability.

\begin{figure*}
\includegraphics[width=1\hsize]{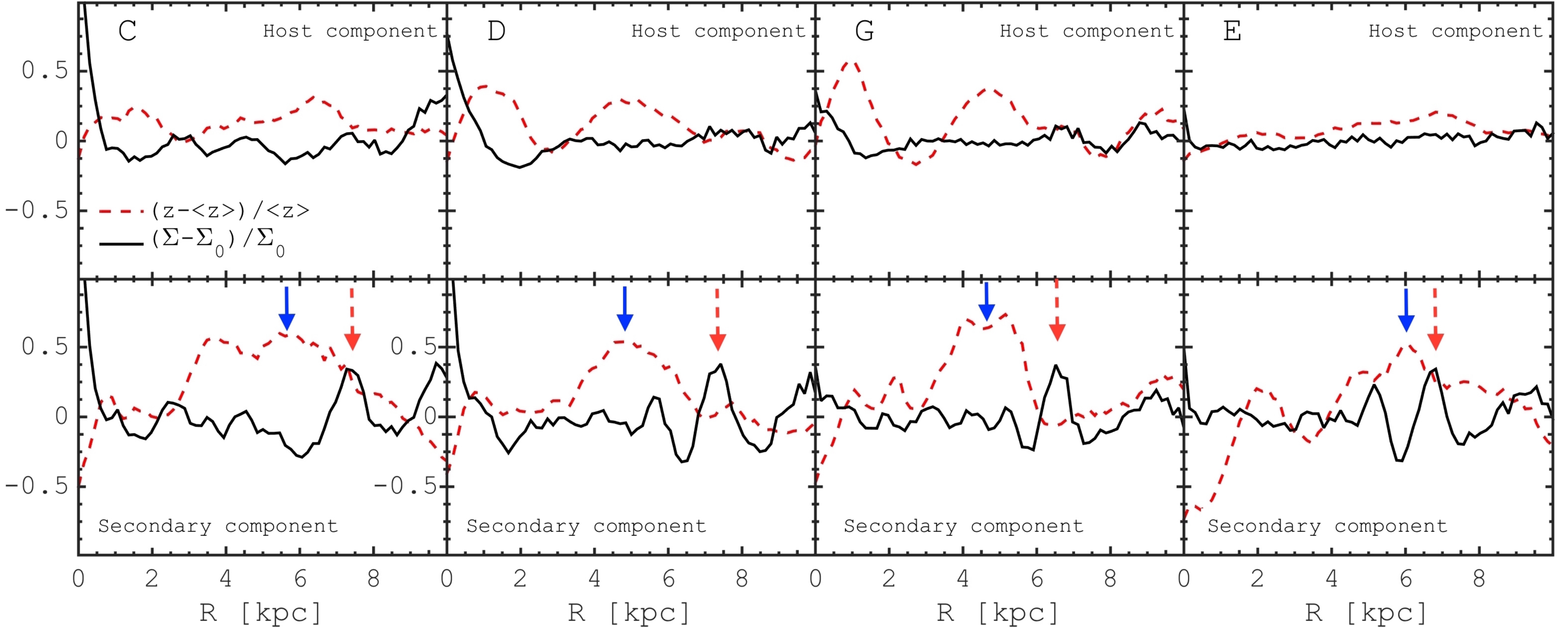}\caption{ Radial profiles of relative perturbations of the disk thickness  (dashed red line) and surface density (solid black line) for the four models with counterrotation~(same models as in Fig.~\ref{fig::rings_faceon}).  Top row shows the data for the host component, bottom row for the secondary component of the simulated galaxy. Blue solid vertical arrows show the position of the bending wave peak. Red dashed vertical arrows show the location of the most prominent ring structure~(see also Fig.~\ref{fig::rings_faceon}). In model C, rings are present, but they are connected with the one-armed spiral structure. Note that in all models the ring structure is strongest in the outer parts of the bending wave.}\label{fig::rings_1d_distr}
\end{figure*}

In our models rings appear to be associated with the bending waves, but actually correspond to density waves in both components~(but seen better in the secondary component). In Fig.~\ref{fig::rings_1d_distr} we plot radial profiles of the disk thickness perturbation and surface density perturbation. For the host component the surface density perturbations are rather small. For the secondary component the maximum of the disk thickness perturbation approximately coincides with the bending wave maximum and for all models there is a clear peak approximately at $R=5-6$~kpc. Note that for the Jeans-unstable model C this peak is less prominent because of the presence of the asymmetric spiral structure. The ring-like structure  is seen clearly as a maximum of the surface density perturbation, and thus it is a {\it density wave}, in all models at $R>6$~kpc. In general, the ring is located well outside the bending wave maximum, where nonlinear effects may be more significant. Thus we conclude that, in our context, stellar rings tend to
form at the outer edge of the bending wave. There is at least one observed case of a detected ring in galaxy with counterrotation. \cite{2013ApJ...769..105K} found a ring-like structure in the K-band brightness distribution of the secondary component of IC~719.

\begin{figure*}
\includegraphics[width=0.5\hsize]{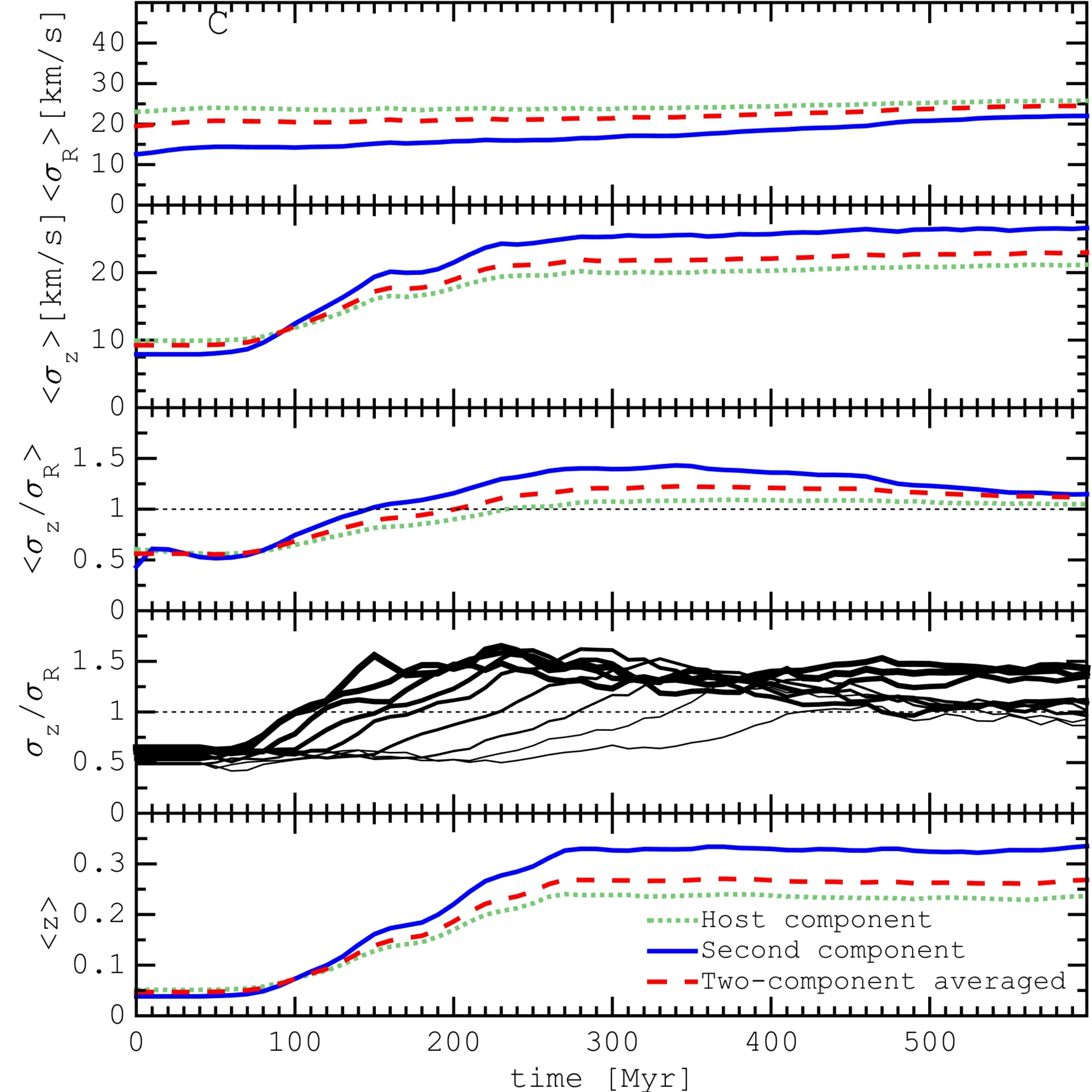}
\includegraphics[width=0.5\hsize]{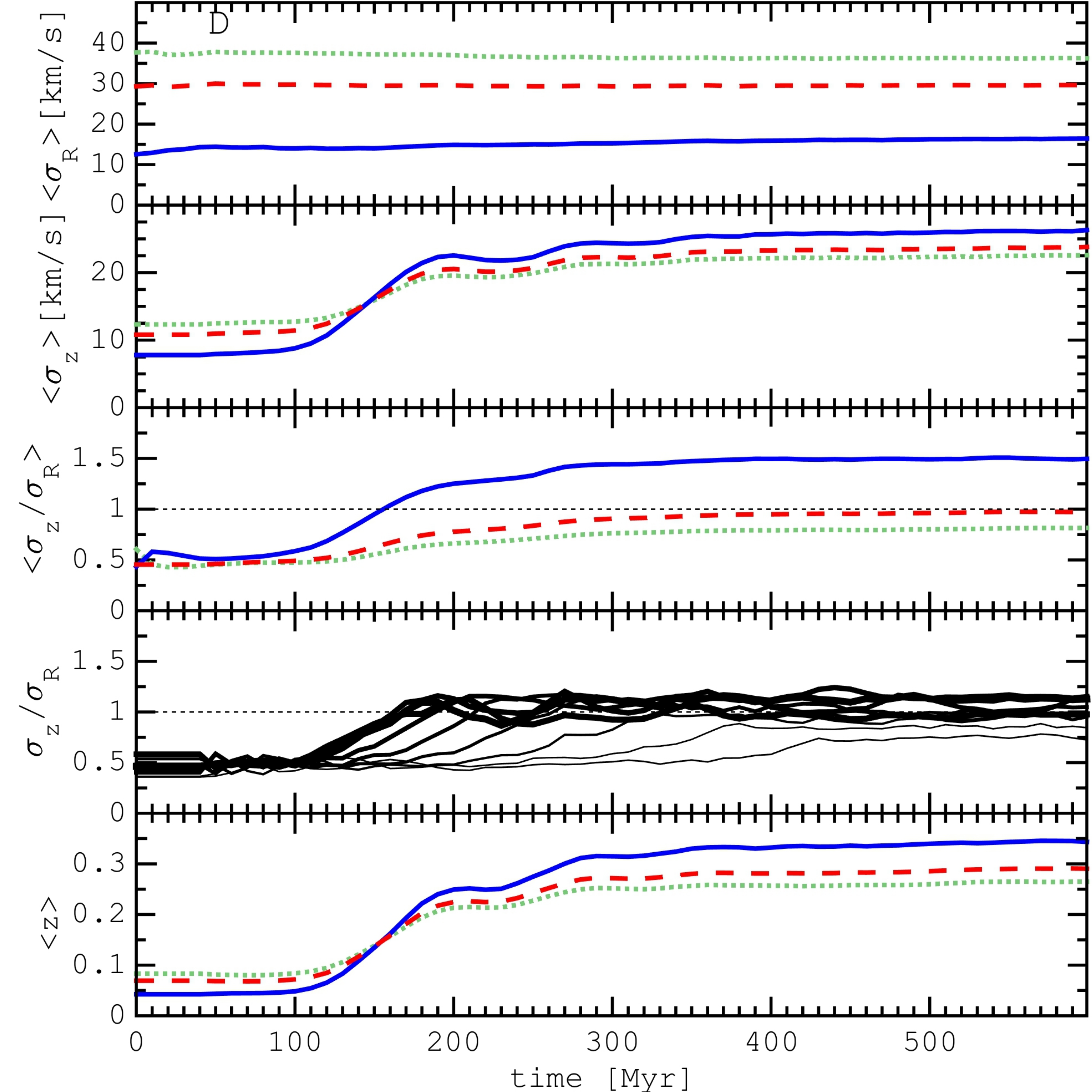} 
\includegraphics[width=0.5\hsize]{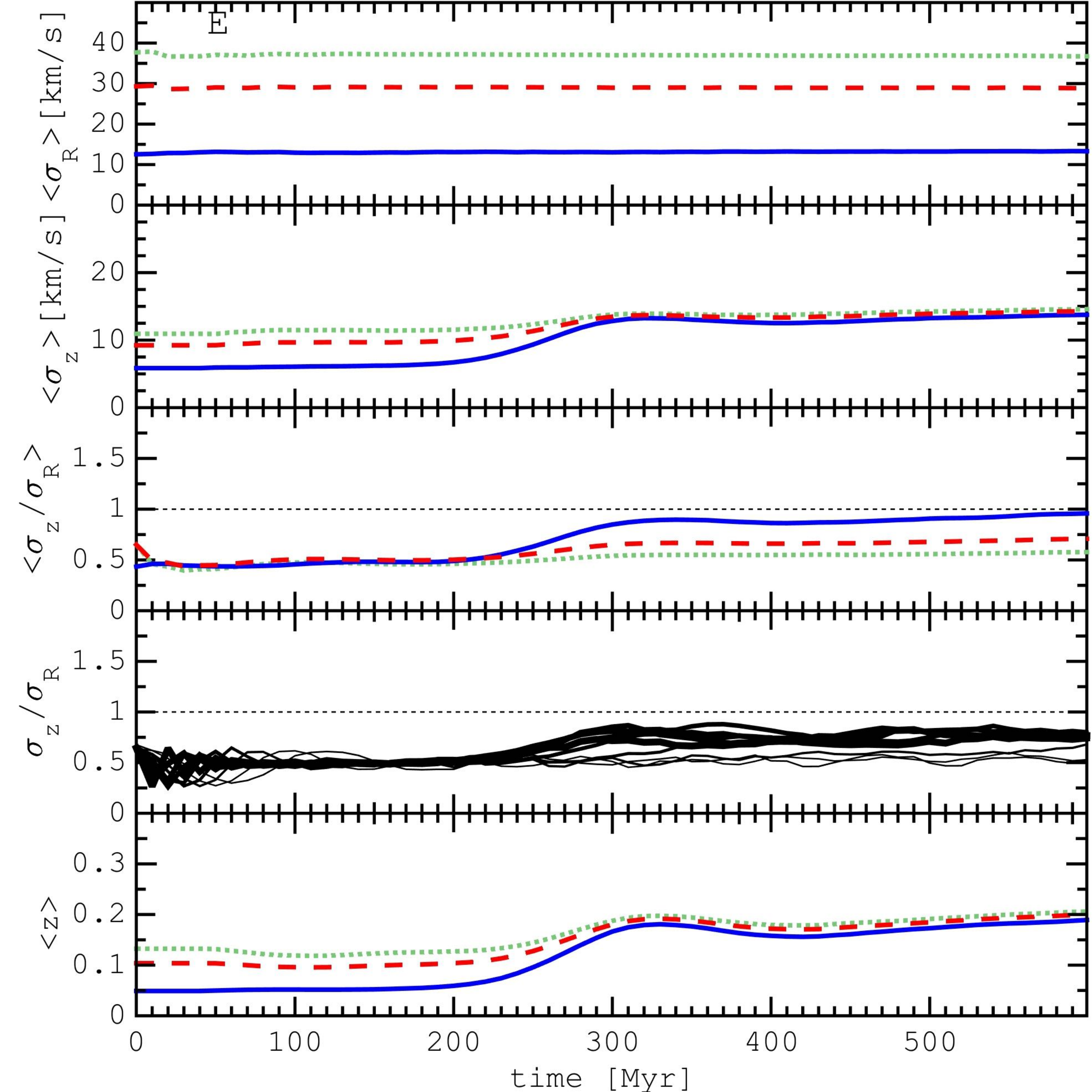}
\includegraphics[width=0.5\hsize]{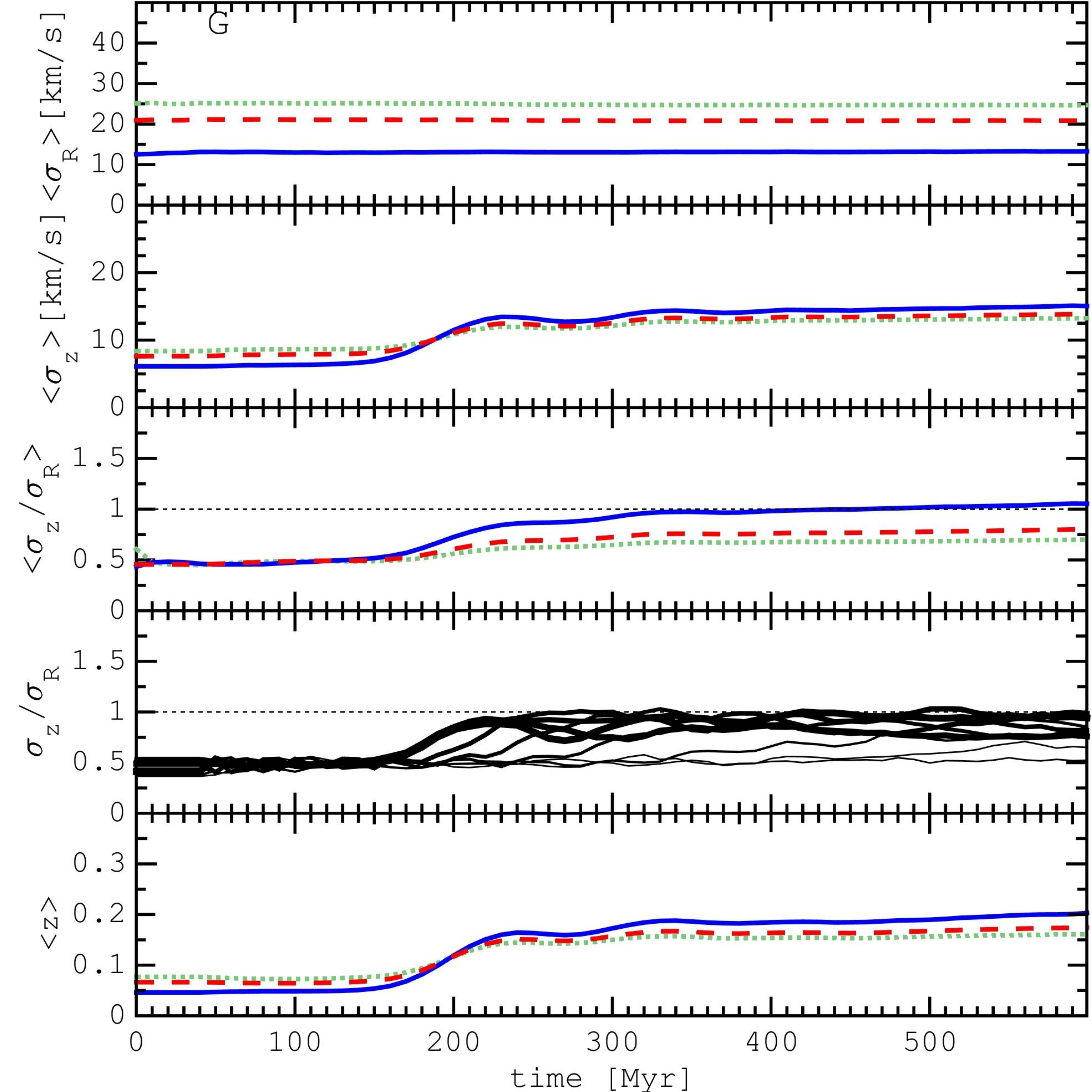} 
\caption{ Evolution of the disk parameters, in the same format as in Fig.~\ref{fig::corot_evol_params}, for four models with counterrotation. Black lines in the fourth frame of each of the four panels show the evolution of two-component averaged values of the velocity dispersion ratio $\czcr$ at various radii (line thickness decreases with  radial distance from the center).}\label{fig::counterrot_evol_params}
\end{figure*}

Figure~\ref{fig::counterrot_evol_params} shows velocity dispersions and disk thickness as a function of time for four simulations with counterrotation. In contrast to the corotating case, there is a clear lack of growth in the radial velocity dispersion$\langle \sigma_{\rm R1}\rangle$ , $\langle\sigma_{\rm R2}\rangle$ (in the disk plane). Only the vertical velocity dispersion increases as a result of strong nonlinear bending instability. As a general trend, this implies an increment of the velocity dispersion ratio $\czcrm$ with time for all models. This growth is quite remarkable, because the dispersion ratio can reach local values from $0.6-0.7$ up to $1-1.5$. We recall that for all models we assume the initial value $\czcr=0.5$ for both components at all radii. In the simulations we find that the constancy of the ratio breaks down for each component, because the radial profile of  $\czcr$ exhibits a significant gradient at the end of the simulations~(see black lines in the third row of each of the four panels of Fig.~\ref{fig::counterrot_evol_params}). Bending waves are known to be able to heat up also one-component galactic disks, but in a very slow process~\citep[e.g., see][]{2013MNRAS.434.2373R}. In contrast, two-component disk relaxation is found to occur in a very short time scale, of about 100-400 Myr.

Bending waves arise initially in the central part of the disk, forming bell-like structures that propagate to the galactic outskirts. While the vertical velocity dispersion increases with time, the radial velocity dispersion remains basically constant for both components. Only unstable model C~($\Qeff(R=R^*)<1$) exhibits a rather slow growth of $\langle \crrr \rangle$ in the counterrotating component (blue circles in the top left frame of Fig.~\ref{fig::counterrot_evol_params}). Such process effectively decreases the velocity dispersion ratio $\langle \czcrt \rangle$ for the less massive counterrotating component, but it does not affect strongly the mean value $\czcrm$ of the velocity dispersion ratio in the entire disk. In conclusion, we find that the vertical dynamics (bending waves and the resulting heating in the vertical direction) is the dominant feature of all models. 

\subsection{General analysis of systems with counterrotation}\label{sec::results_general}
The above results were obtained from simulations in which the disk surface density ratio
is $\alpha = 0.5$; therefore, two thirds of the total disk mass is in the host disk and one third is in the secondary, counterrotating component. To extend the parameter space, we have also run separate simulations with values of the density ratio in the range $\alpha = 0.1-1$ (see third group of models in Table~\ref{tab::tabular1}). Then we have studied two additional models (Kb and Kc), with different initial values of velocity dispersion ratio for the two components (at $R=R^*$, $\czcro=0.7$, $\czcrt=0.4$ for the Kb model and $\czcro=0.6$, $\czcrt=0.2$ for the Kc model; see Table~\ref{tab::tabular1}). In all these models the mean velocity dispersion ratio also increases with time up to $\czcrm = 0.6-1.2$. The stability parameter $\Qeff(R=R^*)$ for these models indicates that the disk is close to conditions of marginal stability with respect to axisymmetric perturbations in the disk plane, but this appears to be unrelated to the vertical heating, which is likely to be associated with the excitation of bending waves.

Since the vertical heating appears to be the most significant feature of our simulations in the presence of counterrotation, regularly present in most of or all the models that we have considered, we tried to see whether the amount of heating observed in the simulations tends to correlate, at the global level, with some properties of the model in its initial state. In addition to properties directly related to the parameters considered in Table~\ref{tab::tabular1}, we decided to estimate the free energy associated with counterrotation in a two-component disk, by introducing the following dimensionless parameter, based on quantities defined in the initial state:
\begin{equation}
\Oo \varepsilon = \frac{\int\limits_0^\infty \pi r \Sigma_2(r) [(r\Omega_2(r))^2 + \sigma^2_2(r)]dr}{\int\limits_0^\infty \pi r \Sigma_1(r) [(r\Omega_1(r))^2 + \sigma^2_1(r)]dr}<1\,,\label{eq::epsilon}
\end{equation}
where $\sigma^2 = \sigma^2_{\rm R} + \sigma^2_{\varphi} + \sigma^2_{\rm z}$. This parameter measures in dimensionless form the strength of counterrotation, in terms of a kinetic energy ratio.

\begin{figure}
\includegraphics[width=1\hsize]{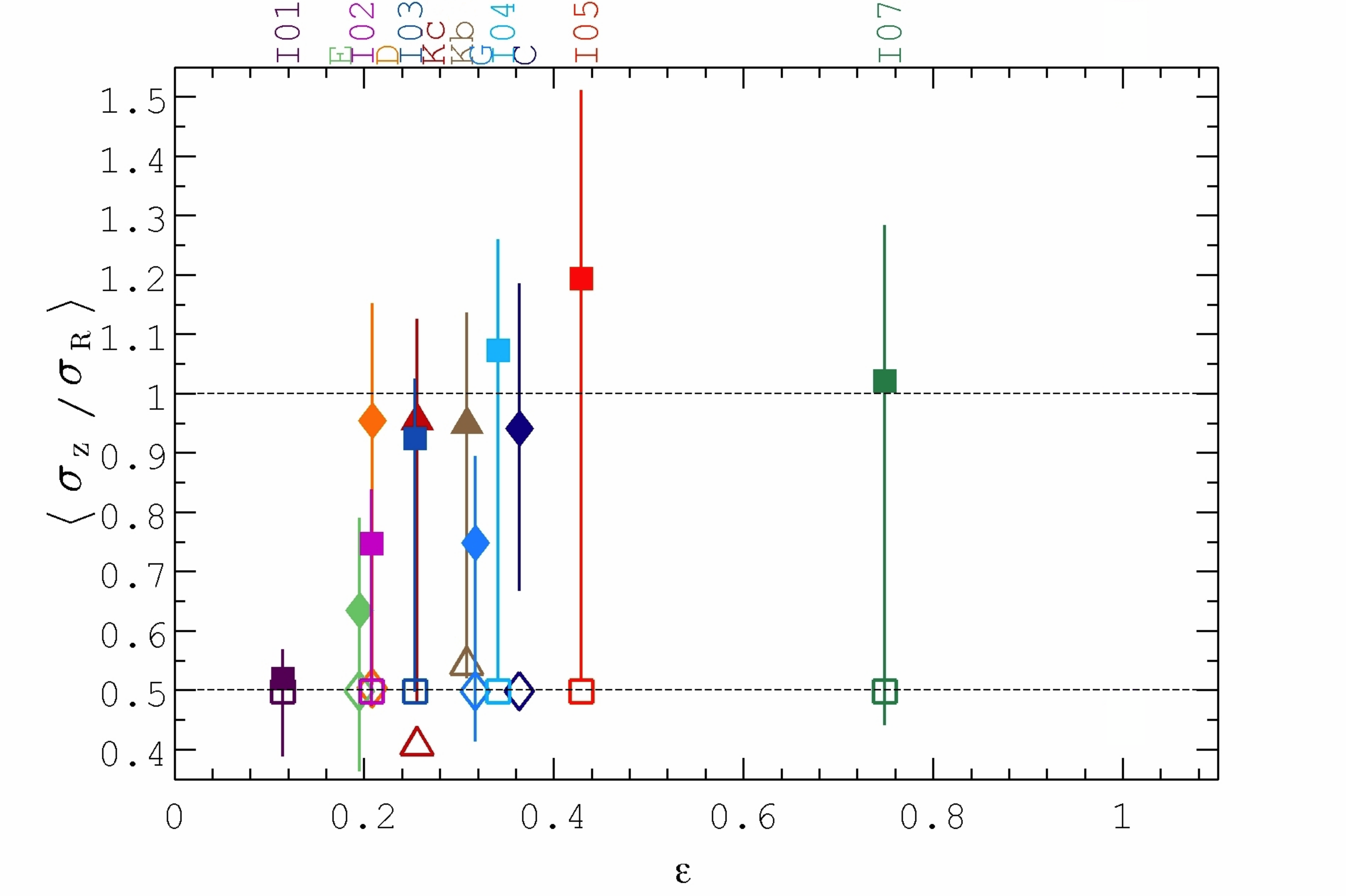}
\caption{ Mean value of the velocity dispersion ratio $\czcrm$ as a function of  $\varepsilon$ at $\approx 600-800$~Myr~(filled symbols) in various models; initial $\czcrm$ values are shown by open symbols. Different symbols correspond to different groups of models in Table~\ref{tab::tabular1}: diamonds for models C, D, E, G; squares for models I01, I02, I03, I04, I05, I07; triangles for models Kb and Kc. Model names are shown above the plot. Vertical error bars correspond to the range of the velocity dispersion ratio, which changes with radius in the final evolved state.}\label{fig::global}
\end{figure}

In Fig.~\ref{fig::global} we plot the evolved mean velocity dispersion ratio $\czcrm$ (at $\approx 600-800$~Myr) as a function of the kinetic energy ratio $\varepsilon$~(see Eq.~(\ref{eq::epsilon})) for all  the models that we have investigated. Figure~\ref{fig::global} shows a reasonable correlation $\varepsilon - \czcrm$, although significant scatter is present:
\begin{equation}
 \czcrm \approx \left\{%
\begin{array}{ll}
    \Oo 1.6\varepsilon + 0.4\,, \varepsilon\leq0.5 \\
    \Oo 1.1\,, \varepsilon>0.5 \\
\end{array}%
\right.
\end{equation}
This interpolating formula covers all the models considered in the present paper, including models Kb and Kc, characterized by initial vertical-to-radial velocity dispersion ratio of 0.55 and 0.4.

\section{Discussion}\label{sec::discuss}
Here we briefly comment on possible consequences that may be relevant to observed galaxies.

We have shown that, especially as a result of thickening and heating induced by bending instabilities, counterrotation can originate significant evolution on a relatively short time scale (200-300 Myr). This evolution is fast and is not to be confused with the results of other secular processes in galaxies~\citep{2006ApJ...645..209D} and slow Jeans-related phenomena that may be difficult to disentangle from numerical relaxation effects~\citep{2013ApJ...769L..24S}. We argue that such fast evolution is expected to take place independently of the formation scenario that may have led to the ``initial" counterrotating system and is likely to hold even in different situations, for example in the case when the secondary counterrotating disk is gaseous; of course, different physical situations would require considering other processes, such as star formation efficiency and depletion time and new simulations should be run to confirm the general features of the evolutionary process or to identify specific phenomena related to the presence of gaseous components.

Our simulations point to final states characterized by high values of the velocity dispersion ratio for both disk components. It would be interesting to gather independent observational constraints on this parameter for individual objects. In turn, this quantity plays an important role in determining the projected line-of-sight velocity dispersion $\sigma_{\rm los}$ and in the disk-halo decomposition of rotation curves.

The acquisition of a counterrotating component may be more frequent than generally believed. In turn, the presence of a secondary counterrotating component at a given epoch may be hard to detect for a number of reasons, not only because the resulting evolution is expected to be fast but especially because a relatively light counterrotating component can easily escape observations. Thus the mechanism of vertical disk heating discussed in this paper may have been overlooked. Certainly, from the theoretical point of view, its occurrence and its impact should be studied further, because in the past little attention has been paid to it.

\section{Conclusions}\label{sec::concl}
In this work we have examined the evolution of two-component collisionless galactic disks by means of numerical simulations. In the presence of counterrotation, bending waves have been found to be excited (two-stream instability). The most prominent and, in a sense, surprising result is that evolution proceeds by effects that change the three-dimensional structure of the system:

\begin{itemize}
\item In the case of counterrotating disks, gravitational instabilities in the disk plane are less effective (with respect to the case of two corotating disks with similar structure). The instability in systems with counterrotation develops mostly in the vertical direction. Strong bending instabilities are excited and lead to a significant increase of the vertical-to-radial velocity dispersion ratio $\czcrm$. In contrast to other relaxation processes in one-component disks, the effects noted in this paper for two-component counterrotating disks occur on the short dynamical time-scale of about 400~Myr.

\item The mean value of the velocity dispersion ratio  in the disk is found to increase up to values close to unity for  both counterrotating components of the disk. The effect is stronger when the free energy associated with counterrotation is larger and then the effects appear to saturate at $\czcrm \approx 1$. For some parameter regimes, in the central regions the velocity dispersion ratio can reach values even higher than unity $\approx 1.2-1.5$~(see Fig.~\ref{fig::global}).

\item The final velocity dispersion ratio is found to change with radius for both counterrotating components~(see Figs.~\ref{fig::counterrot_evol_params},\ref{fig::global}).

\item We found that density-wave ring-like structures are driven by bending instabilities in galaxies with counterrotation. We clearly demonstrate that the outer edge of the bending wave is associated with the rings seen in the face-on density maps.

\item A correlation has been found between the final ``equilibrium" velocity dispersion ratio $\czcrm$ and a dimensionless parameter $\varepsilon$ that we have introduced to characterize the strength of counterrotation. Namely, we found that the $\czcrm$ increases from $0.4$ up to $1.0-1.2$ in the range $\varepsilon = 0-0.5$. At larger values of the counterrotation strength ($>0.5$) the resulting velocity dispersion ratio saturates around unity.

\item We argue that vertical disk heating due to bending waves induced by counterrotation may be a relatively common heating mechanism. Indeed the direct observation of counterrotation may be rare just because the resulting evolution is fast (as shown in this paper) and because relatively light counterrotating disks may escape observations.
\end{itemize}

\begin{acknowledgements}
We wish to thank the Referee for thoughtful suggestions that have improved the quality of the paper. The numerical simulations have been performed at the Research Computing Center~(Moscow State University) under the Russian Science Foundation project~(14-22-00041, ``VOLGA -- A View On the Life of GAlaxies"). This work was partially supported by the President of the RF grant (MK-4536.2015.2), RFBR grants (15-32-21062, 16-02-00649, 16-32-50021) and by the Italian MIUR. SAK has been supported by a postdoctoral fellowship sponsored by the Italian MIUR. SAK also thanks Volgograd State University for hospitality during the final stages of this work.
\end{acknowledgements}

\bibliography{references} 

\end{document}

%% file: notes.tex
\newcommand{\kmps}{km s\ensuremath{^{-1} }\,}
\newcommand{\Msun}{M\ensuremath{_\odot}}
\newcommand{\Msunpc}{M\ensuremath{_\odot} pc\ensuremath{^{-2} }}
\newcommand{\Oo}{\displaystyle}

\newcommand{\czcr}{\sigma_{\rm z}/\sigma_{\rm R}}
\newcommand{\czcro}{\sigma_{\rm z1}/\sigma_{\rm R1}}
\newcommand{\czcrt}{\sigma_{\rm z2}/\sigma_{\rm R2}}
\newcommand{\czcrm}{\langle\sigma_{\rm z}/\sigma_{\rm R}\rangle}
\newcommand{\czcrmo}{\langle\sigma_{\rm z1}/\sigma_{\rm R1}\rangle}

\newcommand{\crrr}{\sigma_{\rm R}}

\newcommand{\Qeff}{Q_{\rm eff}}

\def\ltsima{$\; \buildrel < \over \sim \;$}
\def\simlt{\lower.5ex\hbox{\ltsima}}
\def\gtsima{\; \buildrel > \over \sim \;}
\def\simgt{\lower.5ex\hbox{\gtsima}}